\documentclass[%
reprint,
twoside,
amsmath,amssymb,
aps,
pra,
prb,
rmp,
floatfix,
showkeys, 
]{revtex4-1}

	\usepackage{newtxtext,newtxmath}
    \usepackage{hyperref}
    \hypersetup{colorlinks=true,
                allcolors=blue,   
                filecolor=blue,
                citecolor=blue,
                linkcolor=blue,
                urlcolor=blue,
                menucolor=blue,
                anchorcolor=blue,
                }
	\usepackage{float}
\usepackage{amsmath}
\usepackage{amssymb}
\usepackage{graphicx}
\usepackage{dcolumn}
\usepackage{bm}
\usepackage{array, boldline, makecell, booktabs}

\usepackage[svgnames, table]{xcolor}
\newcommand{\eq}{\ = \ }
\usepackage{appendix}

\begin{document}
	\preprint{}
	\title{
			Stripe patterns orientation resulting from nonuniform 
			forcings \\ and other competitive effects 
			in the Swift-Hohenberg dynamics
	      }
	\author{Daniel L. Coelho}
	\affiliation{GESAR Group/UERJ - State University of Rio de Janeiro,
				 20940-903 Rua Fonseca Teles~121, Rio de Janeiro, RJ, Brazil}
	\author{Eduardo Vitral}
	\affiliation{Department of Mechanical Engineering, University of Nevada,
        1664  N.  Virginia  St.  (0312),  Reno,  NV  89557-0312,  U.S.A.}
	\author{Jos\'e Pontes}%
	\affiliation{GESAR Group/UERJ - State University of Rio de Janeiro,
		         20940-903 Rua Fonseca Teles~121, Rio de Janeiro, RJ, Brazil}
	\author{Norberto Mangiavacchi}
	\affiliation{GESAR Group/UERJ - State University of Rio de Janeiro,
		         20940-903 Rua Fonseca Teles~121, Rio de Janeiro, RJ, Brazil}
	\begin{abstract}
		Spatio-temporal pattern formation in complex systems presents
		rich nonlinear dynamics which leads to the emergence of periodic 
		nonequilibrium structures. One of the
		most prominent equations for the theoretical and numerical study
        of the evolution of these textures is the
		Swift-Hohenberg (SH) equation, which presents a bifurcation
		parameter (forcing) that controls the dynamics by changing the energy 
		landscape of the system, and has been largely employed in phase-field
        models. Though a large part of the literature on pattern
		formation addresses uniformly forced systems,
		nonuniform forcings are also observed in several natural systems, for
		instance, in developmental biology and in soft matter
		applications. In these cases, an orientation effect due to
		forcing gradients is a new factor playing a role in the
		development of patterns, particularly in the class of stripe
		patterns, which we investigate through the nonuniformly forced
		SH dynamics. The present work addresses 
		amplitude instability of stripe textures induced by forcing gradients,
		and the competition between the orientation effect of
		the gradient and other bulk, boundary, and geometric effects
		taking part in the selection of the emerging patterns.
		A weakly nonlinear analysis suggests
		that stripes are stable with respect to small amplitude perturbations
        when aligned with the gradient, and become unstable to such perturbations when
		when aligned perpendicularly to the gradient.
        This analysis is vastly complemented by a numerical work that accounts for other
        effects, confirming that forcing gradients drive stripe alignment, or even
        reorient them from preexisting conditions. However,
		we observe that the orientation effect does not always
		prevail in the face of competing effects, whose hierarchy is suggested
		to depend on the magnitude of the forcing gradient. Other than the
		cubic SH equation (SH3), the quadratic-cubic (SH23) and
		cubic-quintic (SH35) equations are also studied. In the SH23
		case, not only do forcing gradients lead to stripe orientation,
		but also interfere in the transition from hexagonal patterns to
		stripes.
	\end{abstract}
	\maketitle
	
	
	\section{Introduction}\label{sec:introduction}
    The Swift-Hohenberg (SH) equation~\cite{Swift-1977}  is a widely
	adopted mathematical model for describing pattern formation in many
	physical systems presenting symmetry breaking instabilities. A
	classical example where this symmmetry breaking occurs is found in
	the emergence of convection rolls in a thin layer of fluid heated
	from below, for which the distance to the onset of instability is
	given by a control parameter or forcing $\varepsilon$. The SH
	equation with a cubic nonlinearity (SH3) first appeared in the
	framework of B\'enard thermal convection between two ``infinite''
	horizontal surfaces with distinct temperatures. Swift and
	Hohenberg~\cite{Swift-1977} derived an order parameter equation
	from the slow modes dynamics, and addressed its equivalence
	to Brazovskii's model~\cite{brazovskii1975phase}, used for studying 
	the condensation
	of a liquid (disordered phase) to a nonuniform state (periodic).
    Manneville \cite{manneville1983two} derived the SH model by an elimination of the
    vertical dependence of the Oberbeck Boussinesq equations by a
    Galerkin expansion of both velocity and temperature fields, considering
    free slip (stress-free) boundary conditions at the top and bottom of the
    convection cell. A similar reduction of the dynamics was also accomplished for
	reaction-diffusion systems where a similar model could be derived
	with an additional quadratic nonlinearity (SH23)~\cite{walgraefliv}.
	When the nonlinear part of the SH dynamics is a simple polynomial in the order
    parameter (absence of advection and nonpotential terms), it is
    possible to derive such dynamics from the variation of a Lyapunov
    functional, which leads to a relaxational gradient dynamics (and, consequently, stationary patterns) that has been explored in materials science
    and soft matter. This functional is often associated with the system's
	energy, allowing for non-local diffusive dynamics, pattern selection
	and emergence of dissipative structures. In this context, the SH
	equation is also considered as the ``model A'' of periodic 
	systems~\cite{provatas2011}, and, therefore, part of the 
	phase-field theory~\cite{Provatas1,Provatas2}, 
	which originates from
	statistical mechanics principles, and whose goal is to obtain
	governing equations for an order parameter evolution (e.g.
	composition, some microstructural feature); it connects thus
	thermodynamic and kinetic properties with \textit{microstructure}
	via a mathematical formalism~\cite{provatas2005multiscale,
    elder2002modeling,elder2004modeling}.

    An extension of the original Swift-Hohenberg equation (SH3) consists
	in adopting a destabilizing cubic term and in adding a quintic one,
    which effectively changes the energy landscape of the system.
	The resulting quintic equation (SH35) admits the coexistence of
	stable uniform and structured solutions, so that localized patterns
	may exist under a uniform control parameter
	\cite{sakaguchi1996,burke2006localized}. This becomes a desirable
	physical feature in systems that allow for the coexistence of phases
	of distinct symmetry~\cite{vitral2019role}. While localized states
	have been extensively studied through SH35, such states can also
	appear as solutions for SH3 and SH23. The issue for SH3 is that at
	the critical point $\varepsilon = 0$, we have a supercritical
	bifurcation representing the transition from trivial to modulated
	solutions. This means that the amplitude of the stripes emerging at
	$\varepsilon > 0$ increases with $\varepsilon$, and coexistence
	between the stripes and the trivial solution is not possible under
    uniform forcing. However, by letting the control parameter$\varepsilon$ depend on space, localized states become possible by
	varying $\varepsilon$ between values above and below the bifurcation point. 
	In turn, this poses the question of what are the consequences
	of control parameter gradients to pattern selection. While such
	gradients have been known to induce pattern orientation, we here
	propose a comprehensive numerical study of these orientation
	effects, how they affect state localization, and how does gradient
	orientation fare against other competing orientation effects.

	Though widely employed for the study of nonequlibrium pattern
	formation, not many works are found before the turn of the century,
	addressing the orientation effect of the gradients on a structure of
	stripes.	
	One of the pioneers in the subject was Walton
	(1983)\cite{walton1983onset}, who considered the onset of convection
	in the Rayleigh-B\'enard problem, with stress-free upper and lower
	surfaces, and perfect insulating sidewalls. Walton assumed a
	Rayleigh number weakly above the critical value at one of the sidewalls,
	and monotonically decreasing towards the bulk of the convection
	cell. Two cases were identified: if the hotter sidewall was
	maintained sufficiently above the critical temperature a structure of
	stripes perpendicular to the sidewall appears. The result was latter
	confirmed by the theoretical works  of Cross
	(1982)\cite{cross1982boundary,cross1982ingredients} and several
	others~\cite{cross1993,Greenside1984,manneville1995dissipative}.
	Oppositely, if the hotter sidewall was maintained slightly above or
	below the critical temperature, a week structure of stripes parallel
	to the walls emerges. The existence of this weak structure had
	already been noted by the experimental work of Sruljes
	(1979)\cite{sruljes1979}. However, the following step, concerning
	the interaction of the subcritical stripes with a supecritical
	structure, in the case where a positive horizontal gradient of
	temperature is imposed, was not accomplished. One of the purposes of
	the present work consists in analyzing this problem.
    
	At the end of the eighties and the beginning of the nineties a group
	at the Free University of Brussels pointed to the tendency of stripes
	to align to the gradient of the control parameter, and also to
	competing effects that, in many cases, dominate the orientation
	effect created by the gradient~~\cite{pontes1994, pontes2008}.
	Coexistence between hexagons of opposite phases, and between
	hexagons and stripes perpendicular to the gradient of the control
	parameter were identified by Hilali \emph{et al.}
	(1995)\cite{hilali1995}, using a Swift-Hohenberg equation with a
	quadratic term.    Malomed \&
	Nepomnyashchy~(1993)\cite{Malomed_1993} showed that the Lyapunov
	functional associated with a Newell-Whitehead-Segel equation governing
	the evolution of a structure of stripes depends on the angle between
	the stripes and the gradient. These authors showed that the Lyapunov
	potential is minimized whenever the stripes are parallel to the
	gradient and attain a maximum when perpendicular to it.

    In the context of Rayleigh-B\'enard convection, there are also two-dimensional
    generalized SH models that account for non-variational terms, by
    including the coupling between the mean flow and the order parameter
    that represents the vertical velocity~\cite{manneville1983two}. This coupling becomes particularly
    significant for low Prandtl number fluids: for instance, in $CO_2$ gas
    a chaotic state may appear near the onset of convection, characterized
    by a persistent dynamics known as \textit{spiral defect chaos}~\cite{morris1993spiral,vitral2020spiral}. 
    Moreover, the mean flow plays an important
    role in how convection rolls approach the boundaries, and the way they
    interact and compete with gradients of $\varepsilon$ on the alignment of
    rolls is still an open question. For broader phase-field applications,
    the advection of the order parameter by the fluid velocity has also
    appeared in two and three-dimensional models based on SH~\cite{huang2004shear,vitral2021phase}, but the
    interplay between such advection, relevant gradients (e.g. temperature)
    and pattern alignment is yet to be explored.

    More
	recently, Hiscock \& Megason~(2015)\cite{hiscock2015orientation}
	considered the orientation of stripe patterns in biological
	systems, using the Swift-Hohenberg equation in presence of a control parameter gradient. They derived
	Ginzburg-Landau type equations for the amplitudes of a stripped structure presenting
	two perpendicular modes, one of them sharing the gradient's direction. Through an asymptotic analysis, the steady state amplitude of
	both modes was perturbed and it was found that the mode with stripes
	perpendicular to the gradient is unstable, so that the resulting structure is
	comprised of stripes parallel to the gradient. In addition, these authors
	added a reaction term to the Swift-Hohenberg equation and numerically observed pattern
	of stripes perpendicular to the gradient of the control parameter. Only
	periodic boundary conditions (PBC) were considered by the authors. Additionally, orientation of stripe patterns in the presence of control parameter drops has been investigated by Rapp et al.~\cite{rapp2016pattern} for the Brusselator model. Besides the control parameter, a spatially dependent pattern wave number in a Swift-Hohenberg type equation with additional anisotropic terms has been studied by Kaoui et al.~\cite{kaoui2015coexistence}, in the context of wrinkling.
	
	The present paper extends Hiscock \& Megason's work~\cite{hiscock2015orientation}. We initially
    present the SH equation that we numerically integrate 
	and discuss the orientation effect of the gradient of the control
	parameter, in absence of other competing effects.
    We then perform a weakly nonlinear analysis which suggests that a preexisting
    structure of stripes perpendicular to the forcing gradient can become unstable
    when submitted to amplitude perturbations, and eventually replaced by stripes
	in a different orientation. In this theory we derive a pair of
	Newell-Whitehead-Segel equations, from which we propose the amplitude
    instability induced by forcing gradients.
    In sequence, we report the results
	of our numerical simulations for the SH equation with different 
	nonlinearities, in presence of nonuniform forcings. We adopted forces in
	the form of spatial ramps, sinusoidal and gaussian distributions of
	the control parameter. We considered for boundary conditions both generalized Dirichlet (GDBC)
	and periodic (PBC), and explored the cubic (SH3),
	quadratic-cubic (SH23) and cubic-quintic (SH35) forms of the
	Swift-Hohenberg equation. While we investigate orientation effects
	due to local nonlinearities outside the forcing term, there is also
	a recent interest \cite{morgan2014swift} on the nonlocal terms that
	could introduce a new lengthscale into the problem (short or
	long-ranged). These nonlocal nonlinearities significantly modify the
	coefficients of the amplitude equations, and open the opportunity
	for future studies on their effects for pattern selection and
	orientation in two-dimensional systems.
	
	The asymptotic results of Sec.~\ref{sec:ampeqn} are compared with
	the numerical experiments. Due to restriction on finite or
	periodic domain, the gradient faces the competing effect of other
	bulk and the boundaries effects, as well as possible discontinuities 
	of $\varepsilon$ (e.g. ramp in a periodic domain).
	The numerical study also allows us to account for various orientation
    scenarios, while our asymptotic analysis is limited to two different
    configurations. For instance, we can ask if, besides
	modes parallel to the gradient, there are other unstable modes in
	finite or periodic domains, or if modes in \emph{any}
	non-perpendicular direction are unstable. All these questions are
	addressed in the present work.
    
    Competing bulk effects appear in the level of forcing, and in the
	interaction with modes in all directions, either existing in
	pseudo-random initial conditions or generated by the nonlinearity of
	the dynamics. If the forcing is sufficiently high, patterns with a high
	density of defects emerge and dominate the orientation effect of the
	gradient. Another bulk effect that appears in nonuniform forcings is when
	somewhere in the domain $\varepsilon$ locally becomes zero. This situation occurs, for
	example, in configurations of the control parameter resulting in
	domains where both subcritical and supercritical regions coexist. In this
	case, the coherence length of the structure, $\xi \sim
	\varepsilon^{-1/2}$, diverges at $\varepsilon=0$, making the domain
	``short'', in the sense of that critical regions are affected by the
	boundaries, no matter how far they are.	
	Boundary effects consist of the well known tendency of stripes to
	approach supercritical sidewalls perpendicularly to walls
	\cite{cross1982boundary,cross1982ingredients,Greenside1984,cross1993,manneville1995dissipative},
	and of the less known effect of approaching subcritical sidewalls in
	parallel to the
	walls~\cite{walton1982onset,walton1983onset,pontes2008}. We call
	this last one as \emph{subcritical boundary effect} or
	\emph{subcritical effect} for short.
	
	The paper is organized as follows: Sec.~\ref{sec:SH_equation} presents
	the SH equation and the associated Lyapunov functional.
	Sec.~\ref{sec:ampeqn} contains the weakly nonlinear analysis, where
	we address the stability of stripes parallel and perpendicular to
	the gradient, with respect to perturbations in any direction.
	Sec.~\ref{sec:SH3} presents the results of the numerical study with
	the SH3 equation. Sec.~\ref{sec:SH35} addresses the same effects
	with the quadratic-cubic and cubic-quintic equations, SH23 and SH35,
	respectively. The conclusions are presented in
	Sec.~\ref{sec:conclusions}, where we tie the different effects and SH forms
	investigated through energy arguments, and present a hierarchy of effects
	suggested by our findings. We included three appendices describing
	the numerical procedures. 
	Appendix \ref{appendix:cross-sections} provides further results on the numerical amplitude,
	and how our analytic predictions fare in the presence of a nonuniform forcing.
	Appendix \ref{appendix:numerical scheme},
	describes the finite difference scheme adopted for solving the
	Swift-Hohenberg equation and lists the parameters used in the
	simulations. Further details of the procedure are given in our
	recent work~\cite{coelho2020}. Appendix~\ref{appendix:numerical
		scheme2} describes the spectral method employed for obtaining steady
	state amplitude profiles in cross sections of the patterns.
	
	
	\section{The Swift-Hohenberg equation}\label{sec:SH_equation}

	The SH equation has the so-called gradient dynamics, which means
	there is a potential, known as a Lyapunov functional, associated
	with the order parameter field $\psi(\mathbf{x},t)$ [with $\mathbf{x}=(x,y)$] 
	that has the
	property of decreasing monotonically during the
	evolution~\cite{C.I-2002}. It can be derived by using the
	$L^2$-gradient flow of the Lyapunov energy functional:
	\begin{eqnarray}
	\nonumber
	\mathcal{F}[\psi]
	&=&
	\int_\Omega {d\mathbf{x}}\frac{1}{2}
	\left\{
	-{\varepsilon(\mathbf{x})}\psi^2
	+
	{\alpha}
	\left[(\nabla^2+q_0^2)\psi\right]^2
	-\frac{2}{3}\zeta\psi^3+
	\right.
	\\
	&&
	\left.
	\quad\quad\quad\quad
	-\frac{\beta}{2}\psi^4
	+\frac{\gamma}{3}\psi^6
	\right\};
	\label{eq:LF1}
	\\
	\partial_t\mathcal{F}[\psi]
	&=&
	-\int_\Omega {d\mathbf{x}}
	\left(\partial_t\psi\right)^2 \leqslant 0.
	\label{eq:dec_LF}
	\end{eqnarray}
	where $\Omega$ represents the domain whose size is commensurate with
	the length scales of the patterns. We consider a regular square domain $
	\Omega:\{x\in [0,L_x]$, $y \in [0,L_y]\}$ with $L_x = L_y$, since in rectangular
    domains rolls tend to align with the longer sidewalls
    \cite{segel1969distant,daniels1978effect,cross1980effect} (which
    would be an additional competing effect). As discussed, $\varepsilon$
	is a control parameter (forcing) that measures the distance from the
	onset of instability, and it may present a spatial dependency.  
	The constants $\zeta$, $\beta$ and $\gamma$ control the energy structure
	of the system, which describes the energy of phases of distinct symmetry
	and their stability (with implications on the bifurcation diagram).
	The constant $\alpha$ may present different physical interpretations
	(e.g. elastic constant), but it is typically scaled to $\alpha = 1$.
	Equation~\ref{eq:dec_LF} states that the 
	Lyapunov functional monotonically  decreases until steady state is
	reached~\cite{C.I-2002}. By taking the variational derivative of 
	Eq.~\ref{eq:LF1} in $L^2$ norm, the following Swift-Hohenberg
	equation is obtained:
	\begin{align}
	\partial_t\psi \eq&
	- \frac{\delta\mathcal{F}[\psi]}{\delta\psi} \;;
	\\
	\partial_t\psi \eq&
	\varepsilon(\mathbf{x})\psi-\alpha
	\left(\nabla^2+q_0^2\right)^2\psi+
	\zeta\psi^2+\beta\psi^3-\gamma\psi^5 \;.
	\label{SH}
	\end{align}
	
	When setting one or two of the constants $\zeta$, $\beta$ and $\gamma$
    to zero, we obtain one of the SH3, SH23 and SH35 equations. We 
	numerically integrate these equations in the two dimensional square
    domain $\Omega$, using a finite difference scheme with second order
    accuracy in both time and space. The time derivative is discretized
    following the Crank-Nicolson method and spatial derivatives are
    discretized with their appropriate centered difference approximations.
	The scheme preserves the monotonically decreasing nature of the associated
    Lyapunov functional, which we assess during the evolution
    for verification purpose. A brief description of the scheme is given in
    Ref. \cite{coelho2020} and summarized in the
    Appendix~\ref{appendix:numerical scheme}.
    We employ two different kinds of boundary conditions:
    Generalized Dirichlet Boundary Conditions (GDBC), with
    $\psi=\nabla\psi\cdot\mathbf{n}=0$ for $\mathbf{x}\in\partial\Omega$,
    and also Periodic Boundary Conditions (PBC). As in Christov and Pontes
    \cite{C.I-2002}, GDBC is chosen because it assures that the energy evolution
    only depends on its production and conservation in the bulk, not on the boundaries.
    The GDBC specify both the order parameter value and first derivative on the boundaries, where the latter induces
    stripes to align perpendicularly to the boundaries (e.g. anchoring effect at
    a substrate due to surface treatment). PBC is generally employed for large
    systems, where the computational domain $\Omega$ is a representative element
    of the system's behavior and physical boundaries would be far apart.

    In Sec.~\ref{sec:SH3} and \ref{sec:SH35} we present a number of numerical results through nine different figures, each presenting a different set of initial conditions, forcings and equation employed. These various sets are required in order to better isolate the effects that compete with the gradient of the control parameter on the alignment and stability of stripes, allowing us to evaluate their hierarchy. The effects include the forcing type (ramp, sinusoidal functions, gaussian distributions), magnitude of the forcing gradient, geometry of the forcing and compatibility of stripes, stripe reorientation due to amplitude perturbations, boundary conditions, among bulk ones (such as presence of defects) often associated with the previous effects. We are also interested in how the control parameter gradient interfere on the bifurcation diagram, particularly in the transition from hexagons to stripes for SH23, and if the previous conclusions also hold for a subcritical bifurcation in the SH35 case. A summary of these investigations can be found on Table~\ref{tab:table0}.

	\begin{table*}
		\caption{\label{tab:table0}Summary of main numerical results. Initial conditions (IC) are stripes (St), pseudo-random (PR) or square patterns (Sq). Different functions $\varepsilon(\mathbf{x})$ are employed as the forcing: ramp ($\varepsilon_r$), sinusoidal ($\varepsilon_s$) and gaussian distribution ($\varepsilon_g$). Each of the simulation groups investigate how various effects interfere on the orientation and stability of stripes.}
			\begin{tabular}{@{}ccccl@{}} 
				\\\hlineB{2.5}
				\textbf{\quad Figure\quad} & \textbf{Equation} &
				\textbf{IC} 
				& \textbf{\quad Forcing\quad} & \textbf{Investigated effects}
				\\
				\hline
				1 & SH3  & St & $\varepsilon_r$ & Boundary conditions, $\nabla\varepsilon$ magnitude
				\\
				4 & SH3  & St & $\varepsilon_r$ & Amplitude instability, stripe reorientation
				\\
				5 & SH3 & PR & $\varepsilon_r$ & IC bias, $\nabla\varepsilon$ magnitude, subcritical region anchoring
				\\
				6 & SH3  & PR & $\varepsilon_s$ & Oscillating sub/supercritical regions, $\nabla\varepsilon$ magnitude
				\\
				7 & SH3  & PR & $\varepsilon_s$ & Diagonal forcing
				\\
				9 & SH3  & St & $\varepsilon_s$ & Amplitude instability, oscillating sub/supercritical regions
				\\
				10 & SH3  & PR & $\varepsilon_g$ & Forcing geometry and stripe compatibility
				\\
				11 & SH23  & St, PR & $\varepsilon_r$ & $\nabla\varepsilon$ impact on the bifurcation diagram (hexagons, stripes)
				\\
				12 & SH35  & Sq  & $\varepsilon_r$ & $\nabla\varepsilon$ impact on a subcritical bifurcation
				\\
				\hlineB{2.5}
			\end{tabular}
	\end{table*}

	\section{Weakly nonlinear analysis}\label{sec:ampeqn}
	The stability of preexisting patterns can be investigated through the
	amplitude equations derived from the Eq. \ref{SH}. These
	equations describe the motion of the amplitude that envelopes the
	oscillating order parameter $\psi$, and from this coarse-grained
	description we can evaluate how perturbations evolve depending on the
	orientation of the initial pattern and of the gradient of the control
	parameter \cite{manneville1995dissipative,cross1993,hoyle2006pattern}.What
    differ our approach from standard treatments is that we account for the
    spatial dependence of $\varepsilon$, such as in Hiscock \& Megason.
    
	Since the
	derivation of the amplitude equations is relatively similar for SH3, SH23
	and SH35, in terms of scaling and considerations between coefficients,
	we will only provide derivation details for the SH3 case
	($\zeta=\gamma=0$). The SH3 has the form
	\begin{equation}
		\label{eq:sh3}
		\partial_t \psi \;=\; \varepsilon \,\psi 
		- \alpha \, ( \nabla^2 + q_0^2)^2 \psi
		+ \beta \,\psi^3
	\end{equation}
	where $\beta < 0$ and $\alpha > 0$ (generally $\alpha = 1$). 
	
	The solution for a two-dimensional stripe or square pattern can be
	written in terms of a superposition between a sinusoidal function in $x$
	and another in $y$
	\begin{eqnarray*}
		\psi(\mathbf{x},t) &=& A(\mathbf{x},t) \, e^{\, i q_0 x}
		+ B(\mathbf{x},t)\, e^{\, i q_0 y} + c.c.
	\end{eqnarray*}
	where $A$ and $B$ are complex amplitudes. A multiscale expansion is
    performed by introducing  slow variables $\{X,Y,T\}$, which are separated
	from the fast variables $\{x,y,t\}$. The amplitudes are modulated along
	the slow variables as $A(X,Y,T)$ and $B(X,Y,T)$, while oscillation of the
	order parameter in the vicinity of $\mathbf{q}_0$ lie in the scale of
	the fast variables.
	
	Assume we have an initial pattern of perfectly aligned stripes in the $x$
	direction, that is, stripes presenting a wavevector $\mathbf{q}_0 = q_0
	\mathbf{j}$ (unit vector in the $y$ direction). By introducing small
	perturbations to the wavenumber in Eq.~\ref{eq:sh3}, such as $q = q_0 +
	\delta q_y$ and comparing terms, we find that for consistency the slow
	variables should scale as
	\begin{equation*}
		X = \varepsilon^{1/4}x, \quad Y = \varepsilon^{1/2}y, \quad T 
		=\varepsilon t \; .
	\end{equation*}
	From this scaling, we note that derivatives in Eq.~\ref{eq:sh3} should
	follow the chain rule accounting for slow and fast variables, so that
	\begin{equation}
		\label{eq:msdel}
		\partial_x \rightarrow \varepsilon^{1/4} \, \partial_X\;,\quad
		\partial_y \rightarrow \partial_y + \varepsilon^{1/2} \, 
		\partial_Y\;,\quad
		\partial_t \rightarrow \varepsilon \,\partial_T\; .
	\end{equation}
	
	For small $\varepsilon$, the order parameter $\psi$ can be expanded
	about the trivial solution as
	\begin{eqnarray*}
		\psi &=& \varepsilon^{1/2} \, \psi_1 + \varepsilon \, \psi_2 
		+ \varepsilon^{3/2} \, \psi_3 + \dots
	\end{eqnarray*}
	
	By substituting the expanded $\psi$ into Eq.~\ref{eq:sh3} with
	derivatives acting in multiple scales as defined in Eq.~\ref{eq:msdel},
	we are able to collect terms from SH3 in powers of $\varepsilon$. This
	way, we obtain equations from each order of $\varepsilon$ that should be
	independently satisfied, from which we can derive the functions $\psi_i$
	and the equations governing the evolution of $A(X,Y,T)$ and $B(X,Y,T)$. At
	orders $\mathcal{O}(\varepsilon^{1/2})$ and $\mathcal{O}(\varepsilon)$,
	using the notation $L_c = (\partial_y^2+\mathbf{q}_0^2)^2$, we find
	\begin{align*}
		&\mathcal{O}(\varepsilon^{1/2}):\; L_c\psi_1 = 0 
		\\[2mm]
		&\Rightarrow \psi_1
		\;=\; B_{11}\, e^{\, i q_0 y} + c.c. \, ,\; A_{11} \;=\; 0
		\\[2mm]
		&\mathcal{O}(\varepsilon):\; L_c\psi_2+L_c\psi_1 = 0
		\\[2mm]
		&\Rightarrow \psi_2
		\;=\; B_{21}\, e^{\, i q_0 y} + c.c.\,, \; A_{21} \;=\; 0
	\end{align*}
	
	The contribution from the nonlinear cubic term appears starting from the
	next order. This term expands as
	\begin{eqnarray*}
		\psi^3 &=& \varepsilon^{3/2}\psi_1^3 + 3\varepsilon^2\psi_1^2\psi_2
		+3\varepsilon^{5/2}(\psi_1^2\psi_3+\psi_1\psi_2^2) + \dots
	\end{eqnarray*}
	Therefore, at order $\mathcal{O}(\varepsilon^{3/2})$ we find
	\begin{eqnarray}
		\nonumber
		L_c\psi_3 &=& (-\partial_T+1 + 4\alpha q_0^2\partial_Y^2 - 
		\alpha\partial_X^4
		-4i\alpha q_0 \partial_Y\partial_X^2
		\\[2mm]
		&& + 3\beta |B_{11}|^2)B_{11}e^{\,i q_0 y}
		+ (\dots)e^{\,i 3 q_0 y} + c.c.                
		\label{eq:o32}
	\end{eqnarray}
	
	By rewriting Eq.~\ref{eq:o32} as $L_c\psi_3 = \theta$, the solvability
	condition associated to this equation is that $\theta$ must be
	perpendicular to the null space of $L_c^*$: $\theta \perp g \in
	Nu(L_c^*)$. This is the Fredholm's Alternative, the condition under
	which the inner product $(\theta,g) = (\psi,L_c^*g) = 0$ is satisfied,
	and the implication for Eq.~\ref{eq:o32} is that the right-hand side
	must be orthogonal to the eigenfunctions $e^{iq_0y}$, and $e^{-iq_0y}$.
	Therefore, by enforcing solvability we obtain the amplitude equation for
	$B_{11}(X,Y,T)$,
	\begin{eqnarray*}
		\partial_T B_{11} &=& B_{11} + \alpha(2q_0\partial_Y-i\partial_X^2)^2B_{11}
		+3\beta|B_{11}|^2B_{11}\;.
	\end{eqnarray*}
	
	We can similarly find the amplitude equation for $A(X,Y,T)$. Since
	$A_{11}=A_{21}=0$, we need to gather terms up to order
	$\mathcal{O}(\varepsilon^{5/2})$, so that from the Fredholm's
	Alternative we obtain,
	\begin{eqnarray*}
		\partial_T A_{31} &=& A_{31} +\alpha(2q_0\partial_Y-i\partial_X^2)^2A_{31}
		+6\beta|B_{11}|^2A_{31}\;.
	\end{eqnarray*}
	
	In order to rewrite these amplitude equations in terms of the original
	variables, we first note that both amplitudes $A$ and $B$ can be
	expanded in the same form as $\psi$, that is
	\begin{eqnarray*}
		A &=& \varepsilon^{3/2} A_{31} + \varepsilon^{2}A_{41} + \dots \; ,
		\\[2mm]
		B &=& \varepsilon^{1/2} B_{11} + \varepsilon B_{21} + \dots
	\end{eqnarray*}
	Therefore, accounting for the possibility of a control parameter with
	spatial dependence, the pair of amplitude equations for the
	two-dimensional SH3 with stripes perpendicular to the $y$ direction is
	\begin{eqnarray}
		\label{eq:amp-a}
		\partial_t A &=& \varepsilon(\mathbf{x}) A
		+\alpha(2q_0\partial_y-i\partial_x^2)^2A 
		+ 6\beta|B|^2A \;,
		\\[2mm]
		\label{eq:amp-b}
		\partial_t B &=& \varepsilon(\mathbf{x}) B
		+ \alpha(2q_0\partial_y-i\partial_x^2)^2B
		+3\beta|B|^2B\;.
	\end{eqnarray}
	That is, we obtain a system of coupled Newell-Whitehead-Segel (NWS) 
	equations \cite{Malomed_1993}.
	
	The amplitude equations allow us to evaluate a preexisting pattern
	stability in the presence of perturbations, and the role played by
	$\nabla\varepsilon$ in such stability. Note that while $A$ and
	$B$ are complex amplitudes, it can be shown that the phase becomes a constant
	for steady state solutions of parallel stripes~\cite{manneville1995dissipative}, so that we focus
	on the equation for the real part of the amplitude in the following analysis.
	
	Assume $\varepsilon = \varepsilon(x)$ is an increasing ramp in $x$ only,
	and that stripes are initially perpendicular to $y$,
	$\mathbf{q}_0 \perp \nabla\varepsilon$, with a steady state
	amplitude $A_0 = 0$.
    Formally, introducing a spatial dependency in
    $\varepsilon$ holds without interfering in the present derivation
    as long as (i) $\varepsilon(x)$ is a slowly varying function (should only
    act on the amplitude scale, wavenumber much smaller than $q_0$) and
    (ii) insisting that this function is still of order $\varepsilon$.
    In the case of the proposed ramp, the steady real amplitude $B_0$ satisfies
	\begin{equation*}
		\varepsilon(x)B_0-\alpha\partial_x^4B_0+3\beta B_0^3 \;=\; 0 \;.
	\end{equation*}
	Away from the bifurcation point ($\epsilon = 0$), the steady state solution
	is approximately
	\begin{equation*}
		B_0 \;\approx\; \left(\frac{\varepsilon (x)}{-3\beta} \right)^{1/2} \; .
	\end{equation*}
	
	By introducing small perturbations $\delta A$ and $\delta B$ to the
	steady state solutions, we find from Eqs. \ref{eq:amp-a} and
	\ref{eq:amp-b} that these perturbations evolve as
	\begin{eqnarray*}
		\partial_t (\delta A) &=& -\varepsilon (x)\delta A 
		+\alpha(4q_0^2\partial_y^2- \partial_x^4) \delta A
		\\[2mm]
		\partial_t (\delta B) &=& -2\varepsilon (x)\delta B 
		+\alpha(4q_0^2\partial_y^2- \partial_x^4)\delta B.
	\end{eqnarray*}
	
	Since the existence of the solution $(A_0=0,B_0)$ requires $\varepsilon
	> 0$, this implies that when stripes are parallel to~$\nabla\varepsilon$,
	the solution $(0,B_0)$ is stable with respect to small
	perturbations $\delta A$ and $\delta B$.
	
	For the case of stripes perpendicular to the~$\nabla\varepsilon$,
	we keep the ramp $\varepsilon(x)$ in the $x$ direction but change the
	preexisting pattern to stripes aligned in the $y$ direction, so that
	$\mathbf{q}_0 \parallel \nabla\varepsilon$. The
	consequence to Eqs. \ref{eq:amp-a} and \ref{eq:amp-b} is that $A$, $B$
	and space derivatives swap, and the following coupled NWS equations are
	found 
	\begin{eqnarray}
		\label{eq:amp-a2}
		\partial_t A &=& \varepsilon(\mathbf{x}) A
		+\alpha(2q_0\partial_x-i\partial_y^2)^2A
		+ 3\beta|A|^2A \;,
		\\[2mm]
		\label{eq:amp-b2}
		\partial_t B &=& \varepsilon(\mathbf{x}) B
		+\alpha(2q_0\partial_x-i\partial_y^2)^2B
		+6\beta|A|^2B\;.
	\end{eqnarray}
	
	Therefore, now we have a steady state solution $B_0 = 0$ and $A_0 \neq
	0$. Due to the $\partial_x^2A$ derivative in Eq.~\ref{eq:amp-a2}, near
	the Turing instability at $\varepsilon = 0$,  $A_0$ will not behave as
	$(\varepsilon(x)/ -3\beta)^{1/2}$. The $A_0$ solution satisfying
	\begin{eqnarray*}
		\varepsilon(x) A_0 + 4\alpha q_0^2\partial_x^2 A_0 + 3\beta A_0^3 &=&0
	\end{eqnarray*}
	should also satisfy boundary conditions $A_0(x=0) = 0$ and
	$A(x\rightarrow\infty) = (\varepsilon(x)/ -3\beta)^{1/2}$. Note that we
	set the bifurcation point at $x = 0$ for simplicity. For constant
	$\varepsilon$, the solution is of the type $A_0 \sim
	\sqrt{\varepsilon}\tanh(x\sqrt{\varepsilon}/2)$, which behaves as $A_0
	\sim x \varepsilon$ for small $x$ and $\varepsilon$. {\color{black} Since
		for $\varepsilon(x) = c\,x$, where $c$ is a positive constant, there is
		no analytic solution $A_0$, so we assume that $A_0 \sim cx^2$ for
		small $x$.}
	
	For stripes perpendicular to the control parameter gradient, perturbations
	$\delta B$ evolve as
	\begin{eqnarray}
		\label{eq:stability_analysis2}
		\partial_t (\delta B) &=& (\varepsilon(x) + 6\beta |A|^2) \delta B \;.
	\end{eqnarray}
	Taking into account $\varepsilon(x) = cx$, $\beta < 0$ and $A_0 \sim c
	x^2$ for small $x$, we conclude the solution $(A_0,B_0=0)$ is unstable.
	Therefore, while stripes with wavevector $\mathbf{q}_0 \perp
	\nabla\varepsilon$ are stable with respect to small perturbations
	$(\delta A, \delta B)$, we find that stripes with $\mathbf{q}_0 \parallel
	\nabla\varepsilon$ are unstable in the vicinity of the bifurcation point.
    This result does not account for any competitive effect; for instance,
    it may no longer be true close to the boundaries~\cite{bestehorn1992study},
    which we leave as a numerical investigation.
	
	In the following sections, we address to the asymptotic height 
	of the amplitude as
	\begin{eqnarray}
		\label{eq:h2}
		h \;=\; \left(\frac{\varepsilon (x)}{-3\beta} \right)^{1/2} \;.
	\end{eqnarray}
	This quantity, which appears from the weakly nonlinear analysis for
	the two studied orientations of stripes, will be addressed to as an
	analytical result, and used for comparison with the attained steady state
	amplitudes of the numerical results.

	
	\section{Competition between the gradient, 
			 boundary and bulk effects -- SH3}\label{sec:SH3}

		In this section we perform a numerical study of the orientation
		effect due to gradients of the control parameter in presence of
		competing effects, which investigates various initial conditions, forcing
		profiles, and complements the stability analysis
		presented in Sec.~\ref{sec:ampeqn}.
		The study was made through a numerical integration of the SH3 equation,
		using a finite difference semi-implicit time splitting scheme that
		has been previously adopted for Swift-Hohenberg \cite{C.I-1997,coelho2020}
		and other nonlinear parabolic equations \cite{Vitral-2018}. As usual,
		we used the parameters $\alpha=1$, and $q_0=1$. Unless otherwise noted, the
		computational domain consisted of $128\times128$,
		which corresponds to a physical domain of $8\times8$ critical wavelengths 
		considering a grid resolution of 16 points per wavelength. Therefore,
		the grid spacing is $\Delta x\approx 1.016[2\pi/(16q_0)]$ when using GDBC
		and $\Delta x=2\pi/(16q_0)$ when using PBC.
		We employed a time integration scheme of second order accuracy
		using the Crank-Nicolson method, with time step $\Delta t=0.5$ for SH3 
		($\Delta t=0.1$ for SH23/SH35).
		
		The results are organized in four subsections
		\ref{sec:PreexistingPatterns}, \ref{sec:random initial conditions},
		\ref{sec:comparison} and \ref{sec:gaussian}, and further details about the
        scheme and parameters used are summarized in Appendix~\ref{appendix:numerical scheme}. 
		
		Subsection~\ref{sec:PreexistingPatterns} shows the results for
		simulations starting from preexisting stripes, with forcings in the
		form of spatial ramps of $\varepsilon$ along the $x$ direction. Both
		Generalized Dirichlet (GDBC) and Periodic (PBC) boundary conditions
		were considered. Cross sections of selected steady state patterns
		are shown, along with the envelopes obtained as the steady state
		solution of Eqs.~\ref{eq:amp-b}, \ref{eq:amp-a2} 
		and \ref{eq:h}.
		
		Subsection \ref{sec:random initial conditions} presents results
		for simulations starting from pseudo-random initial condition,
		forced with ramps of $\varepsilon$ along the $x$ direction.
		In Secs.~\ref{sec:comparison} and \ref{sec:gaussian} we describe the
		results obtained with sinusoidal and gaussian forcings,	respectively,
		both cases starting from pseudo-random initial conditions.
		
		The patterns presented
		in this work are at the steady state, unless we explicitly state that a specific pattern is transient or at the initial condition $t=0$. We consider a pattern at the steady state when its
		rate of evolution, $\dot L_1$, falls below $5\times10^{-7}$~(see
		Appendix~\ref{sec:L1}). All simulations starting from  pseudo-random
		distribution of $\psi$ used the same distribution as an initial condition.

		\subsection{Spatial ramps of $\varepsilon$ and preexisting 
				    patterns}\label{sec:PreexistingPatterns}
\begin{figure}
		\includegraphics[width=.48\textwidth]{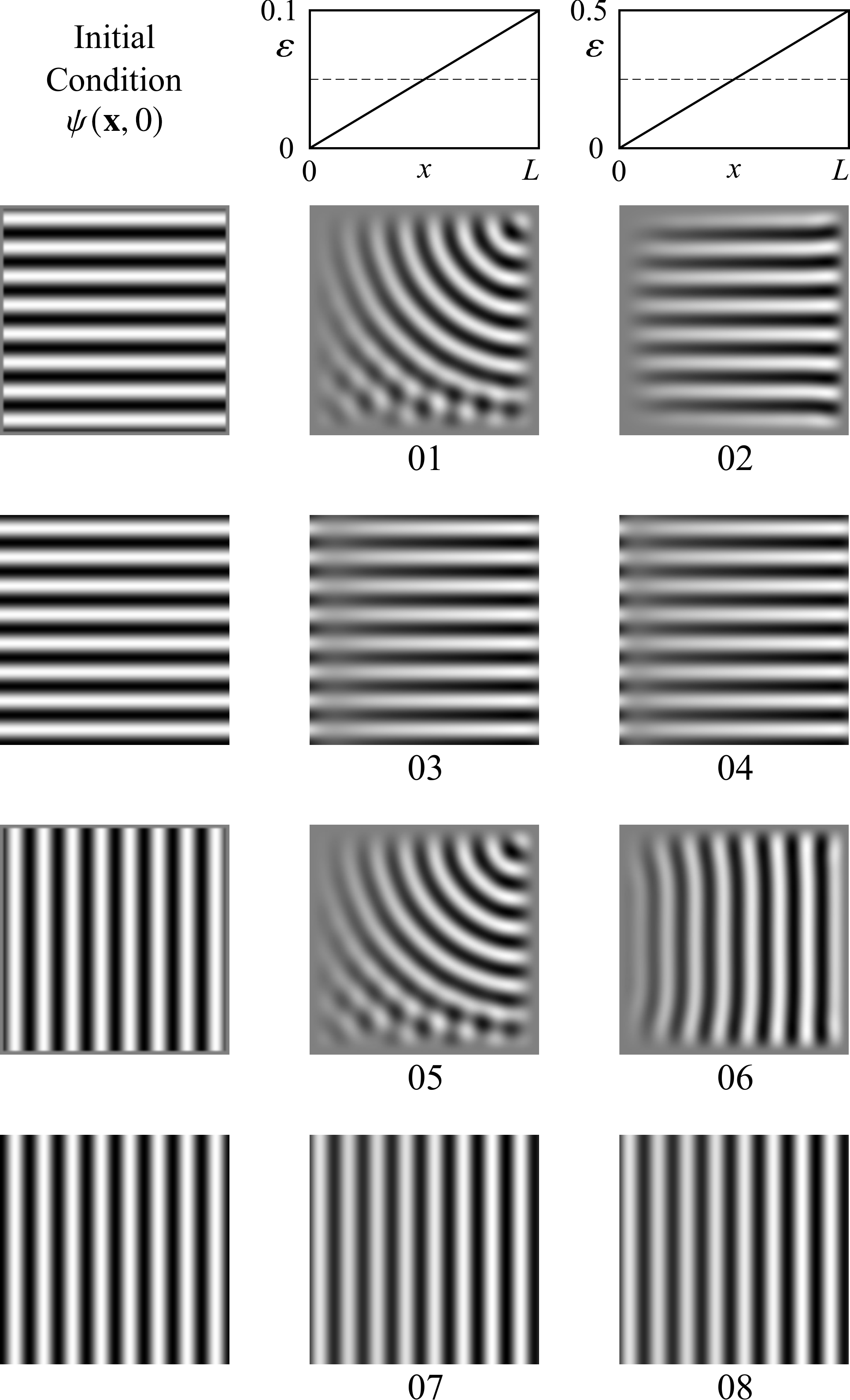}
		\caption{\label{fig:lyaresults1}The results of eight
			simulations with the SH3 equation, preexisting structures, and
			forced with a spatial ramp of the control parameter. The first
			column presents the prescribed initial condition. Columns 2 and
			3 show the attained steady state. The ramp of the control
			parameter is given by the diagrams of the first row. Rows two and
			four correspond to simulations with rigid boundary conditions
			(GDBC), while results presented in rows three and five were
			obtained with periodic boundary conditions (PBC).}
	\end{figure}

\begin{figure*}
	\begin{tabular}{c}
		\centering
		{\includegraphics[width=.99\textwidth]{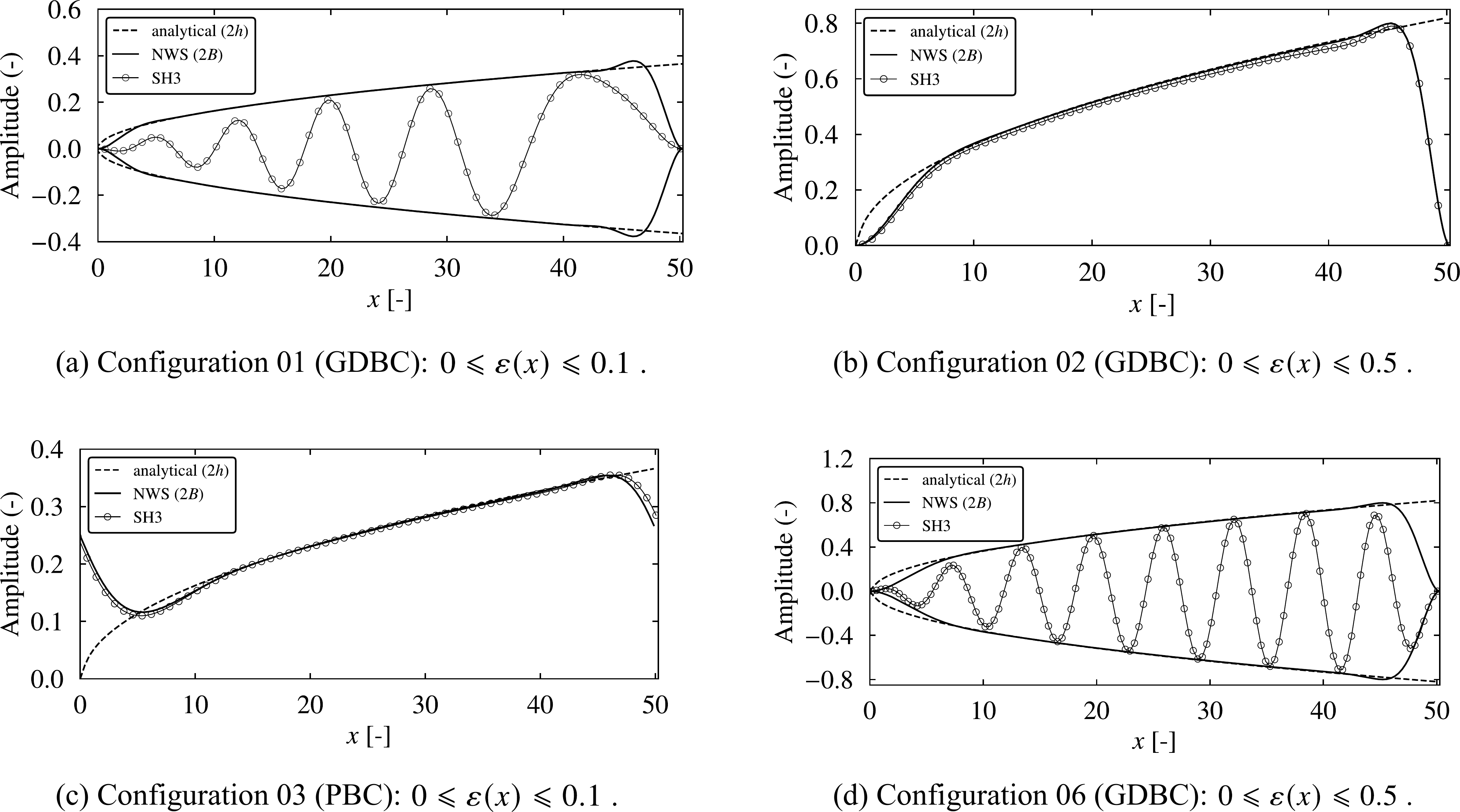}}
	\end{tabular}
		{\caption{\label{fig:comparison2}Cross sections along
		the $x$-direction, taken at the middle of the height of the
		domain ($y$-direction), and envelopes obtained by the two
		methods based on the weakly nonlinear analysis described in
		Sec.~\ref{sec:ampeqn}. The pattern profiles extracted directly
		from the integration of the SH3 equation are represented by
		dotted lines. The envelopes estimated with Eq.~\ref{eq:h} are
		shown in dashed lines, and the envelope obained as the steady
		state solutions of the NWS Eqs.\ref{eq:amp-a} and \ref{eq:amp-b}
		are represented in continuous lines. Eq.~\ref{eq:h} gives a good
		estimation of the envelope, except close to the boundaries,
		since this equation does not take boundary conditions into
		account. Note that the so obtained envelopes in case ($a$) should
		not necessarily fit the crests of the pattern, since these
		curves refer to single mode stripes, either parallel or
		perpendicular to the gradient, and not to bent stripes, where
		the angle of the wavevector with the gradient continuous varies
		across the domain. Nevertheless, the envelope estimated with
		Eq.~\ref{eq:h} matches well the pattern crests, except at the
		boundaries.}}
	\end{figure*}

	This subsection presents results of eight simulations whose
	initial conditions consist of preexisting structures of
	straight stripes parallel or perpendicular to the~$\nabla\varepsilon$. 
	The forcing takes the form of spatial ramps along
	the $x$ direction, submitting the dynamics to this spatial variation
	of $\varepsilon$. The results are shown in
	Figs.~\ref{fig:lyaresults1} and \ref{fig:comparison2}. 
	Figure~\ref{fig:lyaresults1} presents the initial conditions and the steady
	state of the simulations. Figure \ref{fig:comparison2} shows the one
	dimensional profile of four patterns from Fig.~\ref{fig:lyaresults1}, taken
	along the $x$ direction, at the middle height ($y$-direction) of the
	domain. For these profiles, we compared the envelopes of
	modes either parallel or perpendicular to the gradient, to analytic 
	and numerical estimates based on the asymptotic analysis
	detailed in Sec.~\ref{sec:ampeqn}. 
	
	Results shown in the second an third row of
	Fig.~\ref{fig:lyaresults1} were run from an initial condition
	consisting of stripes parallel to the gradient, while
	the last two rows started from stripes perpendicular to the
	gradient. Initial conditions are shown in the first column. GDBC was adopted for the simulations in the second and the
	fourth rows, and those in the third and fifth rows
	were obtained adopting PBC. The preexisting structure of stripes is
	shown in the first column of Fig.~\ref{fig:lyaresults1}. The
	configurations (steady states) shown in the second and third column
	are numbered for reference.
	
	Configuration 1, with a $\varepsilon$ ramp increasing from 0 to 0.1,
	evolved from stripes parallel to the $x$ axis to a
	bent structure of stripes approaching the upper and the right
	sidewalls, oriented perpendicularly to the walls. At the left sidewall, this
	structure is parallel to wall, a result that complies with the work
	of Walton (1982)\cite{walton1982onset}, who identified
	the onset of a weak structure of stripes parallel to a slightly
	subcricrical or supercritical sidewall, in presence of a negative
	gradient of the Rayleigh number pointing to the bulk of a
	Rayleigh-B\'enard cell. A weak structure of stripes perpendicular to
	the lower wall is visible close to that wall,
	since GDBC favors this orientation due to the zero normal derivative.
	Boundary effects dominate both the bulk effects represented by the
	initial condition, and~$\nabla\varepsilon$.
	
	A different situation occurs in the case of configuration~2.
	The gradient, along with orientation of the initial condition, force the
	preexisting structure to remain parallel to it, and stripes are kept
	straightly aligned up to the steady state. This result is attributed
	to the increase in slope and magnitude of the $\varepsilon$ ramp,
	which now increases from 0 up to 0.5.
	
	The cases represented by configurations 3 and 4 of
	Fig.~\ref{fig:lyaresults1} were run with PBC. In both cases, the
	orientation effect of the gradient, along with the initial condition
	and the lack of the competition with boundary effects, result in stripes
	that remained parallel to the gradient, independently of the forcing
	magnitude.
	
	Configuration 5 presents a result similar to the one obtained in
	configuration 1, while in configurations 6, 7 and 8, the
	preexisting initial conditions persist, even with PBC, where
	boundary effects are suppressed.
	The resulting orientation is, in these cases, dominated by
	the preexisting pattern and boundary effects, opposite to the
	orientation favored by $\nabla\varepsilon$. The results of
	configurations 6, 7 and 8 suggest that without a certain level of
	perturbation, the presence of the gradient is not enough to destabilize
	stripes in finite or periodic domains, in the sense that no reorientation
	by the gradient is observed. 
	
\begin{figure}
		\begin{tabular}{c}
			{\includegraphics[width=0.475\textwidth]
			{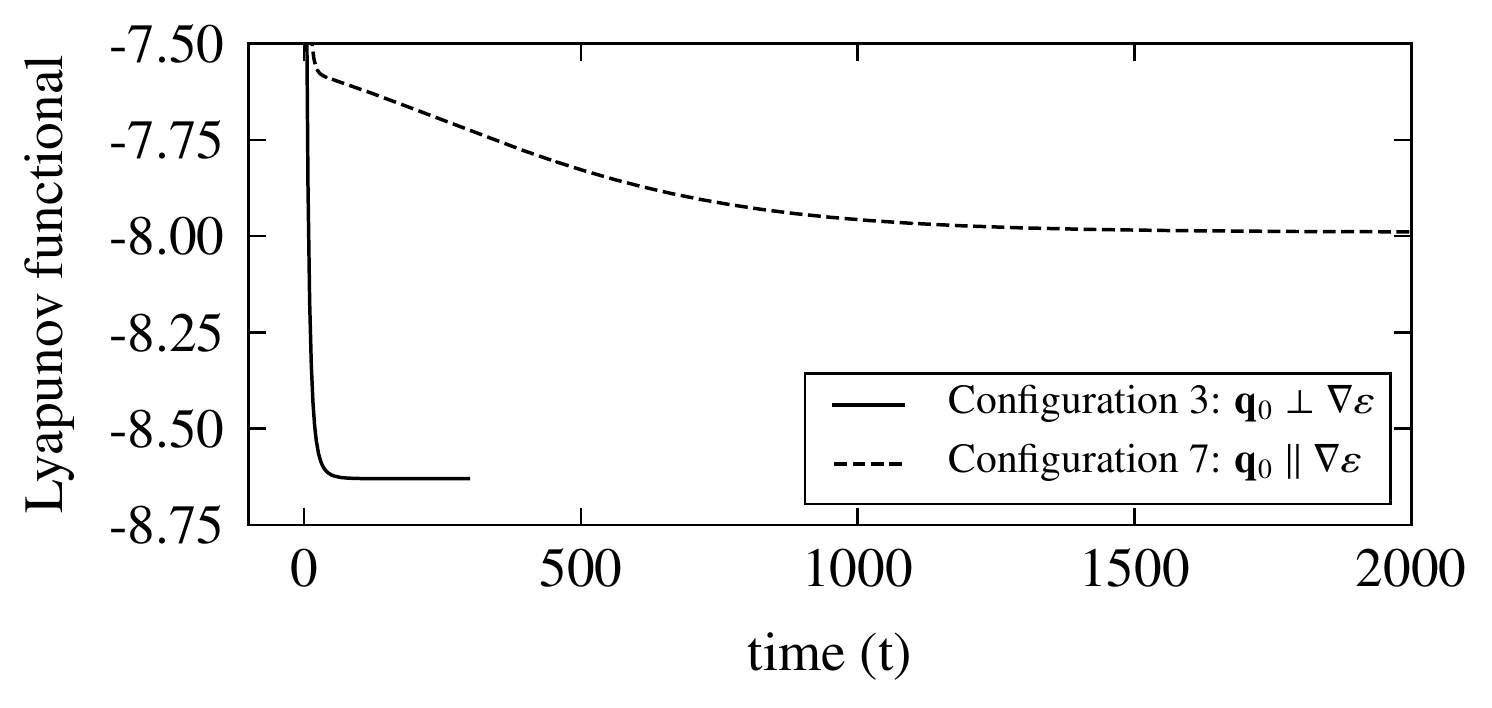}}\\
			{\includegraphics[width=0.475\textwidth]
			{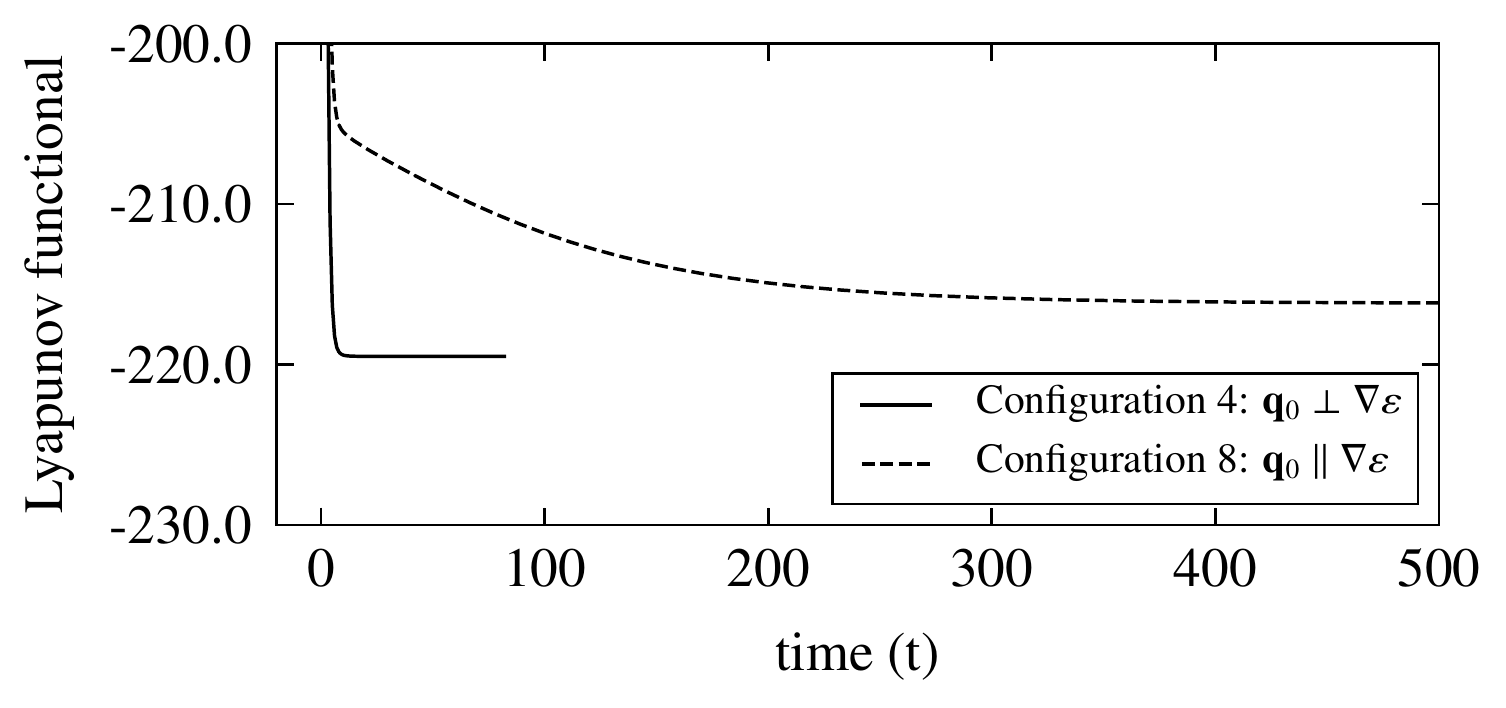}}
 		\end{tabular}
		{\caption{ \label{fig:Lya_curves}Lyapunov functional
		curves of configurations with ramped forcings with each
		configuration indicated. It roughly corresponds to the
		``normalized'' modulus of the time derivative
		$\partial\psi/\partial t$ and therefore is sensitive not only to
		the growth of the amplitude, but also to the pattern phase
		dynamics (Appendix~\ref{appendix:numerical scheme}).}}
	\end{figure}

	While $\nabla\varepsilon$ is not sufficient to reorient a
	preexisting pattern with initial wavevector $\mathbf{q}_0$
	parallel to it (without perturbing the
	system), Fig.~\ref{fig:Lya_curves} shows through the evolution of
	the Lyapunov functional from Eq. \ref{eq:LF1} that the orientation
	of the pattern with respect to the gradient strongly affects the
	relaxational dynamics and energy of the steady state structures. The
	top panel of Fig.~\ref{fig:Lya_curves} follows the energy for the
	dynamics leading to configurations 3 and 7 in
	Fig.~\ref{fig:lyaresults1}, and the bottom panel follows the energy
	evolution leading to the patterns in configurations 4 and 8. For
	both comparisons it is evident that the configuration of stripes
	parallel to the $\nabla\varepsilon$ is the one of minimum energy
	(among the two), whose amplitude quickly relaxes to satisfy the
	control parameter ramp. However, for $\mathbf{q}_0 \parallel
	\nabla\varepsilon$, the relaxation towards the steady state is much
	slower, since this orientation is penalized by the gradient and the
	steady state pattern presents a higher associated energy. 

	We also observe that for higher forcing levels and higher ramp slope
	(bottom panel), the system achieves the steady state faster than for
	lower forcing levels (top panel). Moreover, the energy ratio between
	$\mathbf{q}_0 \perp \nabla\varepsilon$ and $\mathbf{q}_0 \parallel
	\nabla\varepsilon$ decreases from $1.080$ between the two steady states
	in the top panel to $1.015$ in the bottom panel. The previous two
	observations suggest that the orientational effects due to
	$\nabla\varepsilon$ weaken as the forcing level increases.
	
	Figure~\ref{fig:comparison2} shows cross sections of configurations 1,
	2, 3 and 6 of Fig.~\ref{fig:lyaresults1}. The cross sections were
	taken along the $x$ direction, at the middle height ($y$-direction)
	of the domain. The acquired profiles were superposed with the
	patterns envelope, estimated using two approaches, based on the results
	from Sec.~\ref{sec:ampeqn}. The first approach
	consists of an asymptotic form $h$ of the envelope height away from the bifurcation
    point, 
    \begin{eqnarray}
	\label{eq:h}
	h \;=\; \left(\frac{\varepsilon (x)}{-3\beta} \right)^{1/2} \;,
	\end{eqnarray}
    which is obtained from the steady state solution of the NWS equation,
    Eq.~\ref{eq:amp-b}, and depends only
	parametrically on $x$ in this approximation. Note that no
	subcritical solutions are possible with this equation. The second
	approach consists in solving the NWS Eq.~\ref{eq:amp-b} at the steady
	state with a pseudo-spectral method described in
	Appendix~\ref{appendix:numerical scheme2}.
	
	For case ($a$) of Fig.~\ref{fig:comparison2} we acquired the profile from
	configuration 1 of Fig.~\ref{fig:lyaresults1}, using GDBC.
	Strictly speaking, the envelopes for this case
	should not necessarily fit the crests of the pattern, since these
	envelopes refer to single mode stripes, either parallel or
	perpendicular to the gradient, and not to bent stripes, where the
	angle of the wavevector with the gradient continuously varies across
	the domain. Nevertheless, the envelope estimated with Eq.~\ref{eq:h}
	matches well the pattern crests, except at the boundaries. The
	amplitude obtained numerically as the steady solution of Eq.~\ref{eq:amp-b}
	matches well the crests of the pattern. This satisfactory agreement
	suggests that the local amplitude depends, in this case, majorly on the
	local value of $\varepsilon$, and not on the local orientation of
	the wavevector.
	
	\begin{figure*}
	\begin{tabular}{c}
		\centering
		{\includegraphics[width=0.825\textwidth]{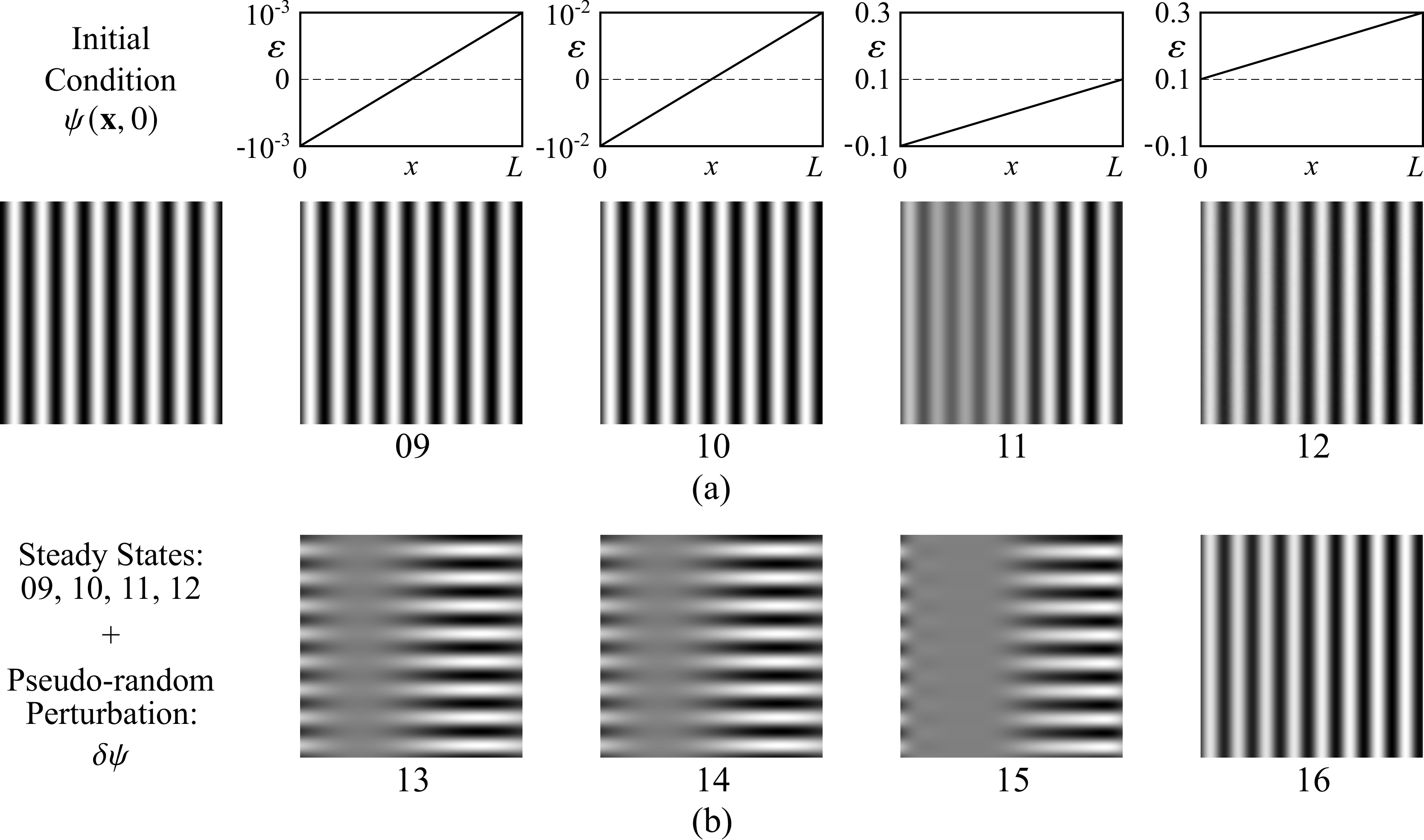}}
	\end{tabular}
	\label{fig:preexisting}
	{\caption{
		The results of eight simulations with the SH3 equation, preexisting
		structures, forced with a spatial ramp of the control parameter and
		PBC. The ramp of the control parameter is given by the diagrams of
		first row. In the first row are shown (a) steady patterns obtained
		from the indicated initial condition of preexisting vertical stripes
		without pseudo-random  perturbation. In the second row are shown (b)
		Steady patterns obtained from the steady states above with the exact
		same pseudo-random perturbation $\delta\psi \in (-10^{-4},10^{-4})$. 
		Periodic domains (PBC)  were
		chosen in order to observe ``only'' interactions between bulk and 
		$\nabla\varepsilon$ effects without the rigid imposition on the
		boundary (GDBC). Since in the weakly nonlinear analysis we find that
		vertical stripes are unstable in the presence of the
		$\nabla\varepsilon$, a small perturbation was sufficient to the system
		evolve into a new steady pattern where stripes align to the
		$\nabla\varepsilon$ direction (Configurations 13 and 15).
		Configuration 14 does not present such behavior and the preexisting
		stripes orientation prevails.}}
	\end{figure*}

	Cases ($b$) and ($c$) of Fig.~\ref{fig:comparison2} refer to
	structures of stripes $B$ parallel to the gradient, using GDBC
	(configuration 02) and PBC (configuration 03), respectively. The cross
	sections capture the amplitude of the stripes along the direction of the
	gradient. The envelope given by Eq.~\ref{eq:h} fits well the amplitude
	of the pattern away from the boundaries. Moreover, an excellent matching
	exists between the envelope obtained as the numerical solution of the NWS
	Eq.~\ref{eq:amp-b}, and the one extracted directly from the pattern,
	with both boundary conditions.
	Finally, case ($d$) of Fig.~\ref{fig:comparison2} was obtained from a
	preexisting structure of stripes $A$, perpendicular to the gradient
	(configuration 06). As in case ($a$), the crests of the pattern fit well
	the envelopes, and reinforces the observation that away from the boundaries
	$\nabla\varepsilon$ dictates the behavior of the amplitude.
	
	In order to evaluate possible reorientation effects due to $\nabla\varepsilon$,
	as suggested by Sec. \ref{sec:ampeqn}, we perturb the steady state configuration
	that originally had a monomodal pattern with $\mathbf{q}_0\parallel\nabla\varepsilon$.
	The orientation effect can be seen in some of the configurations from
	Fig.~\ref{fig:preexisting}, using PBC, so that we avoid effects from the rigid
	imposition on the boundary in the GDBC case. From a monomodal initial condition,
	we first obtain steady states under different ramps of $\varepsilon$, as seen in 
	configurations 9, 10, 11, and 12, which preserve the original preexisting vertical
	stripes. By imposing a pseudo-random perturbation of $\delta\psi \in (-10^{-4},10^{-4})$
	to these steady states, the cases with a ramp of $\varepsilon$ crossing the bifurcation point
	at $\varepsilon = 0$ reorient into horizontal stripes, as observed in
	configurations 13, 14, and 15, independently of the slope and magnitude
	used for the ramp in $\varepsilon$. This result agrees with the suggestion from the
	stability analysis (Eq. \ref{eq:stability_analysis2}), as in face of an amplitude
    perturbation, stripes in the neighborhood of the bifurcation point are expected to
	become unstable when $\mathbf{q}_0\parallel\nabla\varepsilon$, leading the pattern 
	to reorient. However, when the
	$\varepsilon$ ramp solely stays in the supercritical regime, $\varepsilon > 0$,
	the initial pattern did not reorient, as shown in configuration 16.
	
		\subsection{Spatial ramps of $\varepsilon$ and 
			        pseudo-random initial conditions}\label{sec:random initial conditions}

\begin{figure}
	\begin{tabular}{c}
		\centering
		{\includegraphics[width=.48\textwidth]{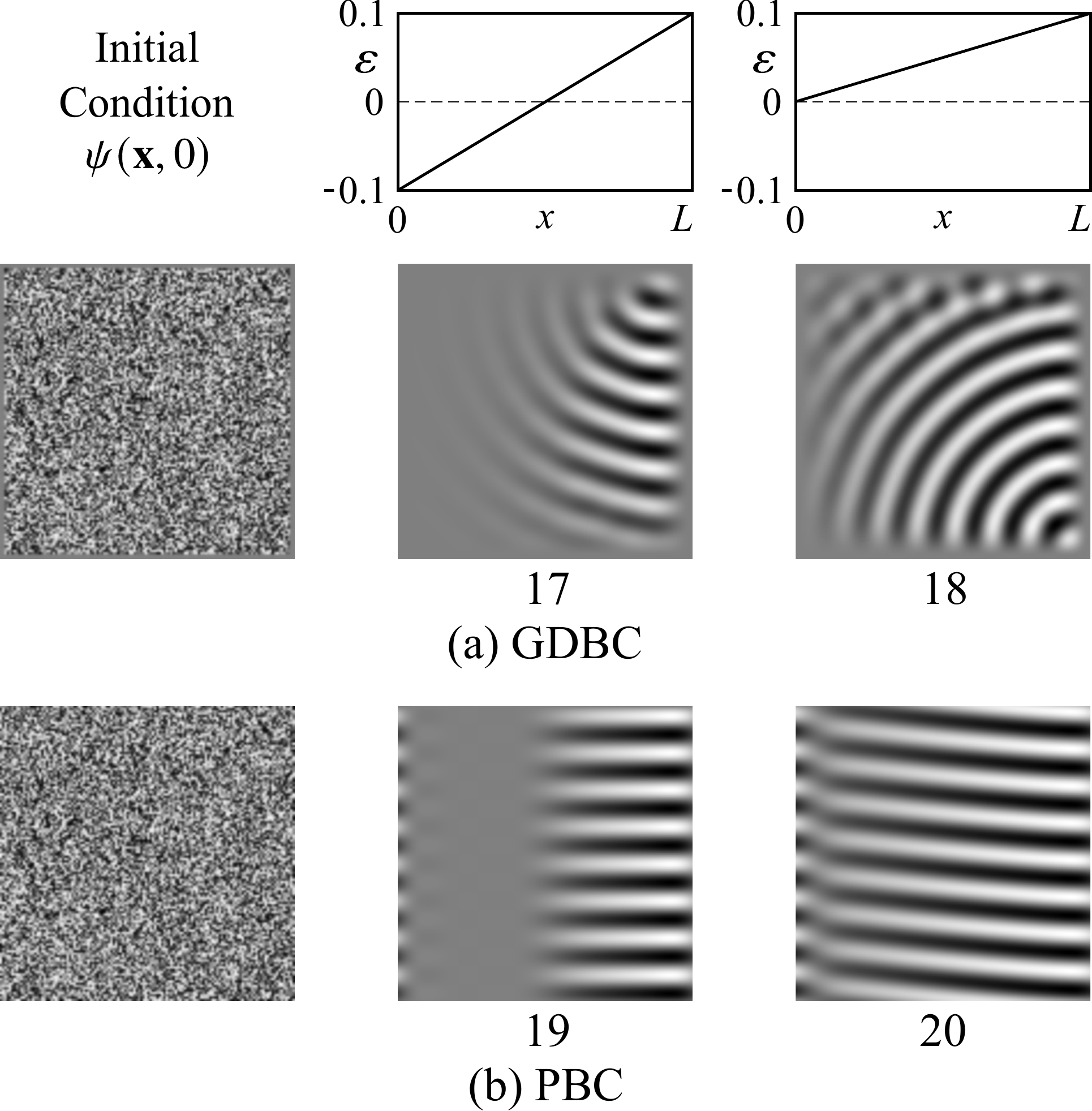}}
	\end{tabular}
    {\caption{\label{fig:lyaresults2}
		The results of four
		simulations with the SH3 equation, forced with a spatial ramp of
		the control parameter $\varepsilon$. All simulations started
		from the same pseudo-random initial conditions, shown in the
		first column. The remaining columns present the steady state.
		First row: the prescribed profile of the control parameter
		$\varepsilon$. Second and third rows: GDBC, and PBC boundary
		conditions, respectively. Simulations of the second row show
		that boundary effects dominate the orientation effect when GDBC
		are prescribed. In the third row, the absence of boundary
		effects allows the dominance of the orientation effect of the
		gradient.}}
	\end{figure}

    In this subsection we present results of four simulations run
	from the same pseudo-random initial condition, and the cross section
	for two of the obtained configurations. Two simulations were performed
	with GDBC, and the other two with PBC. The results are sumarized in
	Fig.~\ref{fig:lyaresults2}.
	
	In the case of configuration 17, the orientation effect of the
	gradient is dominated by the boundary and the subcritical effects:
	a bent structure of stripes perpendicular to the right and to the
	upper sidewalls emerges. This structure persists in the subcritical
	region, with weak stripes approaching the left sidewall in parallel
	orientation. The result is in agreement  with the works of
	Sruljes~(1979)\cite{sruljes1979} and Walton~(1982)\cite{walton1982onset}.
	
	The steady state pattern
	developed in configuration 17 is similar to the one appearing
	in configuration 1 of Fig.~\ref{fig:lyaresults1}, with the
	orientation effect of the gradient dominated by the boundary and
	the subcritical effects. The structure developed in configuration 18
	is similar to the one of configuration 17, with stripes
	perpendicular to the supercritical lower and right sidewalls, and
	parallel to the critical left sidewall. Additionally an weak
	structure of stripes perpendicular to the upper supercritical
	sidewall is visible.
	
	In the case of PBC, configurations 19 and 20, no boundary effects
	are present and the resulting structure selects a direction almost
	parallel to the the gradient, even with the presence of modes in every
	directions in the initial condition. A Benjamin-Feir instability
	appears at the left limit of the periodic structure of
	configuration 20.

		\subsection{Sinusoidal forcings}\label{sec:comparison}

\begin{figure*}
	{\centering
	 \includegraphics[width=0.99\textwidth]{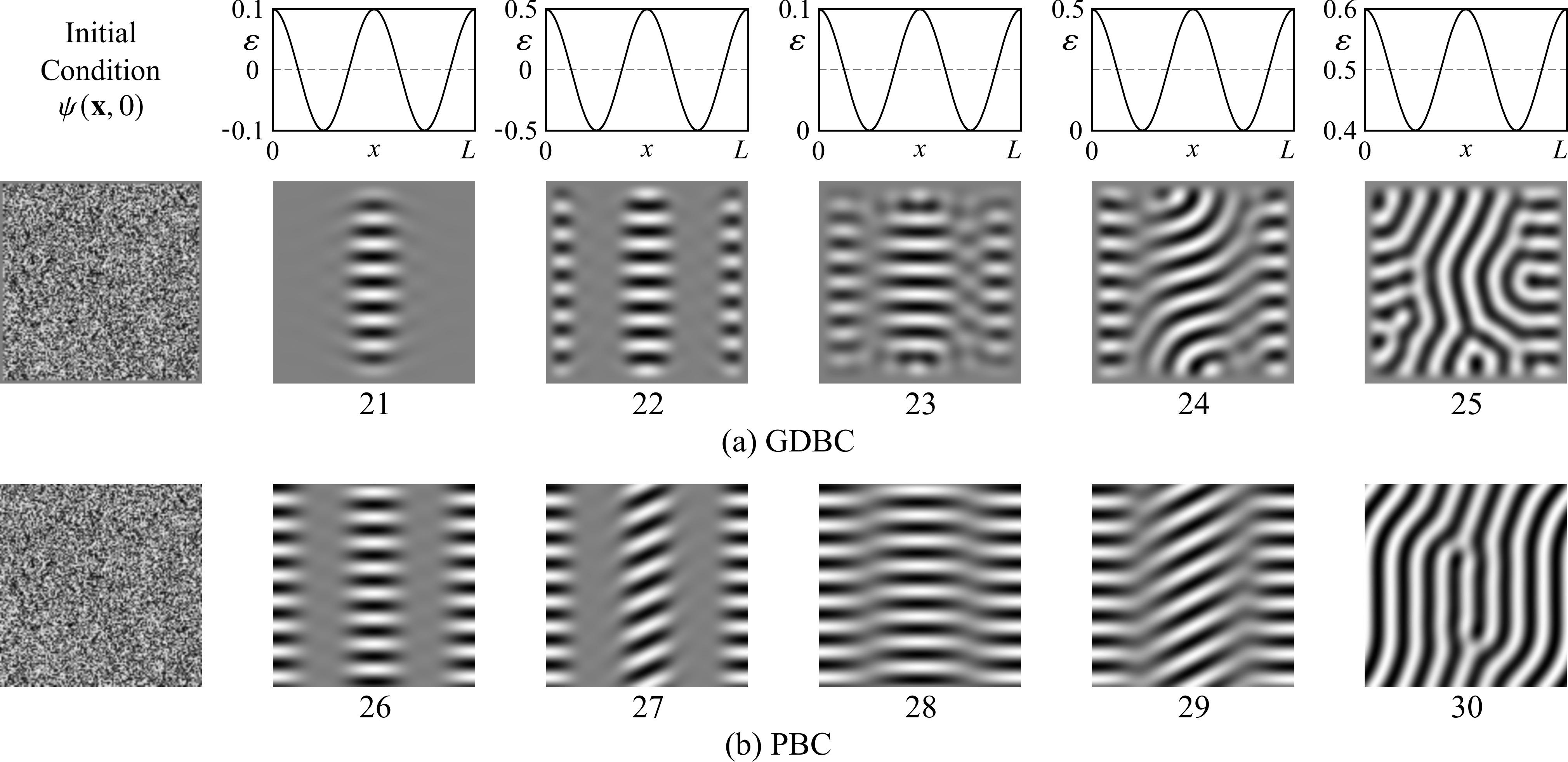}}
	\caption{\label{fig:lyaresults3}The results of ten
			simulations with the SH3 equation, forced with a spatial
			sinusoidal profile of the control parameter~$\varepsilon$. All
			simulations started from the same pseudo-random initial
			conditions, shown in the first column. The remaining columns
			present the steady state, attained when $\dot L_1\leq 5
			\times10^{-7}$. First row: the 	prescribed profile of the
			control parameter $\varepsilon$. Second and third rows: GDBC
			(a), and PBC (b) boundary conditions, respectively.}
	\end{figure*}

	This subsection presents results of eight simulations run from
	pseudo-random initial conditions with sinusoidal forcings in $x$. Both
	GDBC and PBC were prescribed. Sinusoidal forcings are of interest
	because they allow for multiple subcritical and supercritical regions
	in a single domain, with multiple bifurcation points $\varepsilon=0$ in $x$. Also, when
	compared to ramps, periodic forcings better accommodate PBC, so that
	undesired effects due to a jump of $\varepsilon$ at the boundary are
	not an issue. The results are summarized in Fig.~\ref{fig:lyaresults3},
	where we present the resulting steady state patterns (labeled as
	configurations 13-22). 
	The first row displays the distribution of $\varepsilon$ forcings, the second row displays the patterns obtained by prescribing GDBC, and the third row displays the patterns obtained by prescribing PBC.
	
	Results shown in configurations 21 to 25 (GDBC) and 26 to 30 (PBC)  of
	Fig.~\ref{fig:lyaresults3} evolved either to patterns parallel
	to the gradient, or at least, with regions where the stripes are parallel
	to the gradient. The orientation effect clearly appears in these
	simulations, even when the restrictive GDBC are prescribed.
	
	To obtain configuration 21, we used a low amplitude sinusoidal
	distribution of $\varepsilon$ with subcritical regions
	($-0.1 \leq \varepsilon \leq 0.1$). Due to GDBC, we observe the
	existence of supercritical regions close to the right and left
	walls where no pattern emerges. A higher amplitude of the forcing,
	as depicted in configuration 30, leads to the emergence of stripes
	aligned to the gradient in all supercritical regions, so that a
	higher $\varepsilon$ allows to overcome energy penalization due to
	boundary conditions. Configuration 23 of Fig.~\ref{fig:lyaresults3}
	was run with a sinusoidal forcing added to a constant, so that 
	$\varepsilon\geqslant0$. No subcritical regions are present in this simulation. A
	weak structure of small stripes perpendicular to the upper and lower walls
	emerges at the center of these walls. Several Benjamin-Feir
	instabilities are also observed. Configuration 24 was run with a similar
	sinusoidal forcing, but of higher amplitude. Due to the  higher forcing,
	the structure can more easily accommodate defects. A pattern of winding
	stripes appear at the central region, with two focus defects showing at
	the top and at the bottom of this region. We note that the winding
	form of these stripes comes from the fact that they anchor perpendicularly to
	the upper and lower walls, while approaching regions of $\varepsilon$
	close to zero with a parallel alignment. This is in agreement with our
	previous observation from Fig.~\ref{fig:Lya_curves}, that orientational
	effects due to $\nabla\varepsilon$ become less prevailing as the magnitude of
	the forcing increases. 
	
	Configuration 25 is obtained by increasing the minimum $\varepsilon$ even
	the positive value of $\varepsilon = 0.4$, so that we have a sinusoidal forcing of higher magnitude and GDBC, while decreasing $\nabla\varepsilon$ magnitude. 
	The orientational effect of the gradient is fully dominated by the bulk and
	boundary effects. A pattern of mostly upwards stripes with a high density of
	defects emerges.
	
	The results presented in the third row of Fig.~\ref{fig:lyaresults3}
	were run with PBC. In the case of configuration 26, a low forcing
	and the lack of boundaries prevent the emergence of defects. The
	orientational effect of the gradient prevails and a structure of stripes
	parallel to the gradient emerges in all supercritical regions.
	Configuration 27 is similar to 26, but grows from a sinusoidal
	forcing of higher amplitude. This higher forcing in supercritical
	regions allows for stripes that deviate from the gradient alignment,
	and we observe columns of stripes that alternate between parallel and
	inclined alignments. The steady state for this case strongly depends
	on the initial distribution of $\psi$, and once a column of
	inclined stripes is formed, it is unable to completely reorient
	in the gradient direction.
	
	Configuration 28 of Fig.~\ref{fig:lyaresults3} presents again a case
	where the forcing consists of a low amplitude with zero minimum
	$\varepsilon$. The absence of sidewalls and the relatively low forcing
	weakens competing effects and pattern aligns accordingly to the gradient.
	The resulting structure is parallel to the gradient and
	Benjamin-Feir instabilities are observed at the neighborhood of the bifurcation point. For configuration 29 we use the same forcing as in configuration 24.
	The steady state pattern is similar to configuration 27, but with
	stripes occupying the entire domain, as the forcing is non-negative.
	Lastly, configuration 30 starts from the same forcing as configuration 25,
	and the resulting winding pattern shows that, even for PBC, $\nabla\varepsilon$
	fails to orient the stripes whenever the magnitude of the force 
	remains at large magnitude.

    \begin{figure}
	\centering
	{\includegraphics[width=0.475\textwidth]{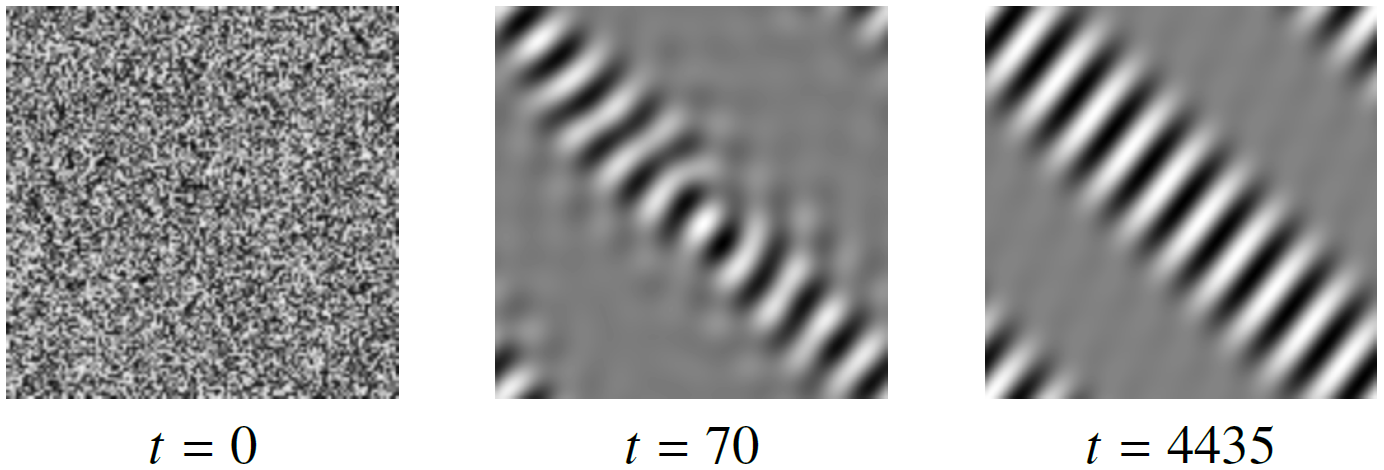}}
		{\caption{\label{fig:45 degrees vertical stripes}
		Pattern evolution from pseudo-random initial conditions, subjected to
		PBC and a diagonal sinusoidal distribution of the control parameter,
		given by: $\varepsilon(\mathbf{x})=0.1\cos{[q_1 (x+y)]}$. The
		structure evolves into stripes aligned in the $\nabla\varepsilon$
		direction until the steady state is reached.}}
	\end{figure}

	Fig.~\ref{fig:45 degrees vertical stripes} presents the result of a
	configuration consisting of a pseudo-random distribution of
	$\psi$ as the initial condition, a sinusoidal forcing along
	the domain diagonal, and PBC. Lack of boundary effects along with the
	orientation effect of the gradient, and existence of modes along all
	direction in the initial condition lead to a pattern of stripes
	parallel to the diagonal. 
	
\begin{figure}
	\centering
	{\includegraphics[width=0.475\textwidth]{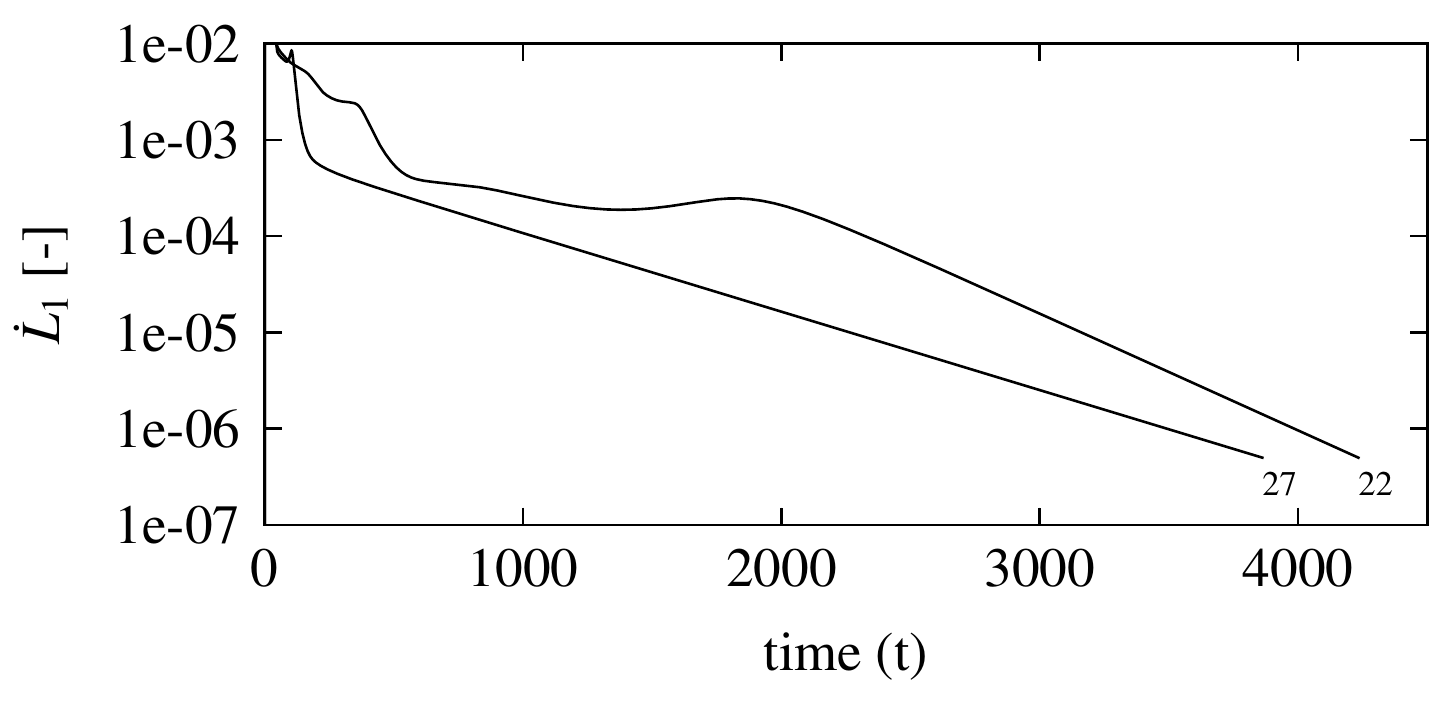}}
	{\caption{\label{fig:L1_curves} 
		Comparison between $\dot{L}_1$ curves of configurations 22 (GDBC) and
		27 (PBC) with sinusoidal forcings. It roughly corresponds to the
		``normalized'' modulus of the time derivative $\partial\psi/\partial
		t$ and therefore  is sensitive not only to the growth of the
		amplitude, but also to the pattern phase dynamics
		(Appendix~\ref{appendix:numerical scheme}).}}
	\end{figure}

	Figure~\ref{fig:L1_curves} shows the $\dot L_1\times t$ curves of
	selected configurations shown in Fig.~\ref{fig:lyaresults3}. These
	curves present an irregular region at the very beginning of the
	simulations, when the patterns emerges from the pseudo-random initial
	condition. Most of the pattern growth occurs at this
	phase. As a result, $\dot L_1$ decreases by some orders of
	magnitude. The evolution proceeds with changes in the phase, and
	with the amplitude essentially saturated. $\dot L_1$ evolves
	irregularly at much lower level, with peaks occurring at the
	collapse of defects. This phase is followed, in all cases, by a
	linear (exponential) decrease of $\dot L_1$. We assume that the
	pattern reached a steady state when $\dot L_1$ attains the value
	$5\times10^{-7}$. We mention that also the Lyapunov potential
	decreases exponentially at this phase.
	
\begin{figure}
	\begin{tabular}{c}
		{\includegraphics[width=0.475\textwidth]{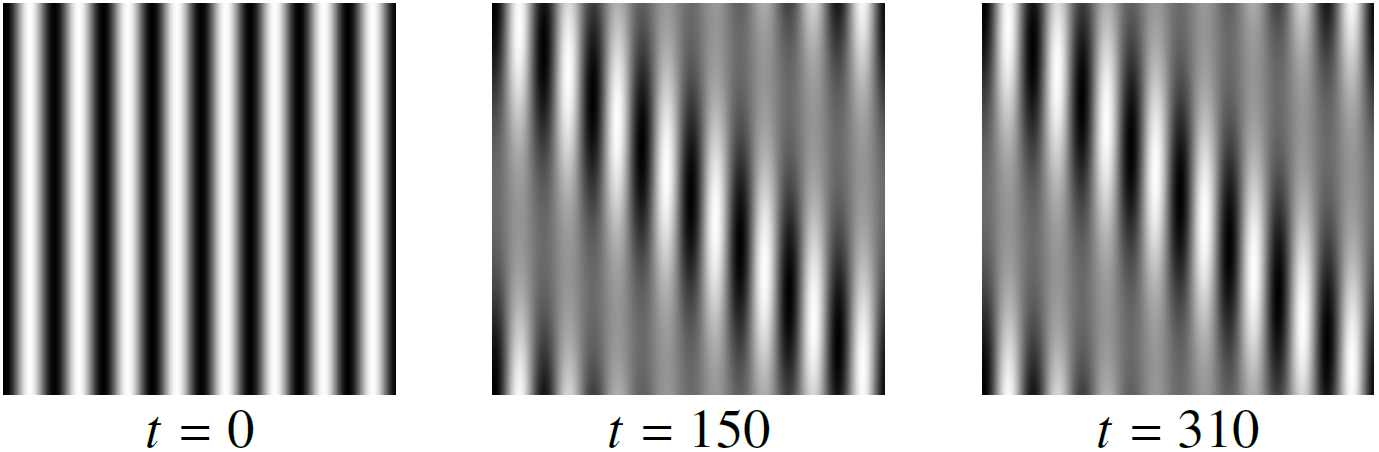}}\\[00pt]
		{(a)}\\[10pt]
		{\includegraphics[width=0.475\textwidth]{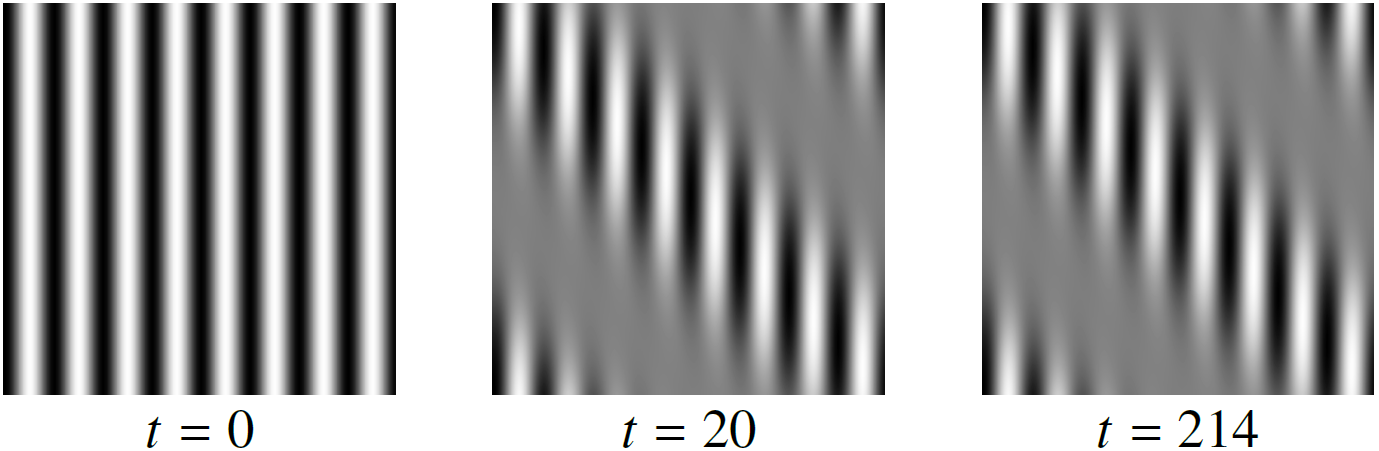}}\\[00pt]
		{(b)}\\[10pt]
		{\includegraphics[width=0.475\textwidth]{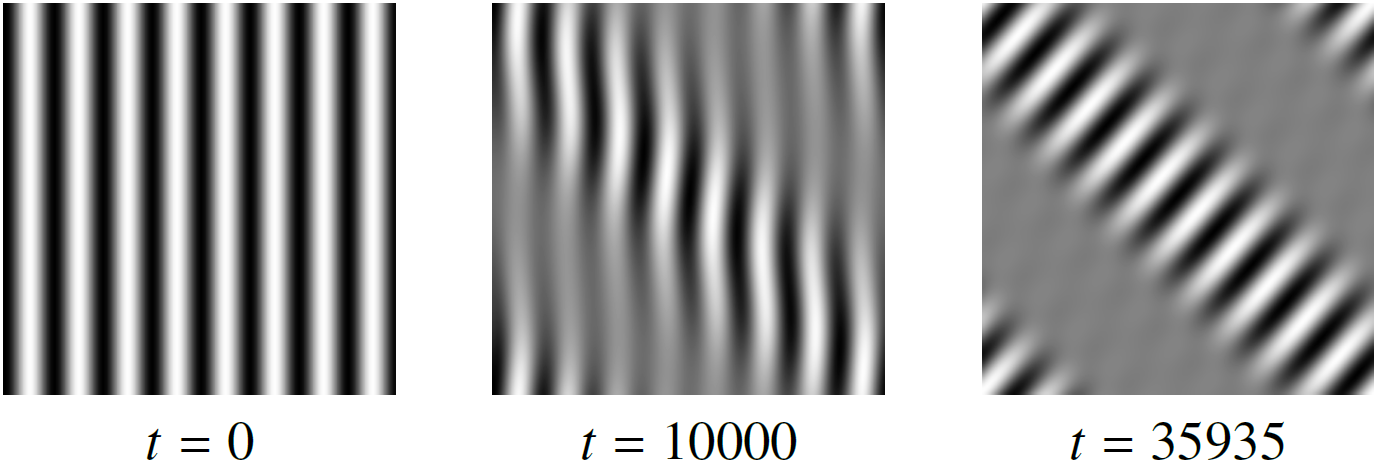}}\\[00pt]
		{(c)}\\[10pt]
		{\includegraphics[width=0.475\textwidth]{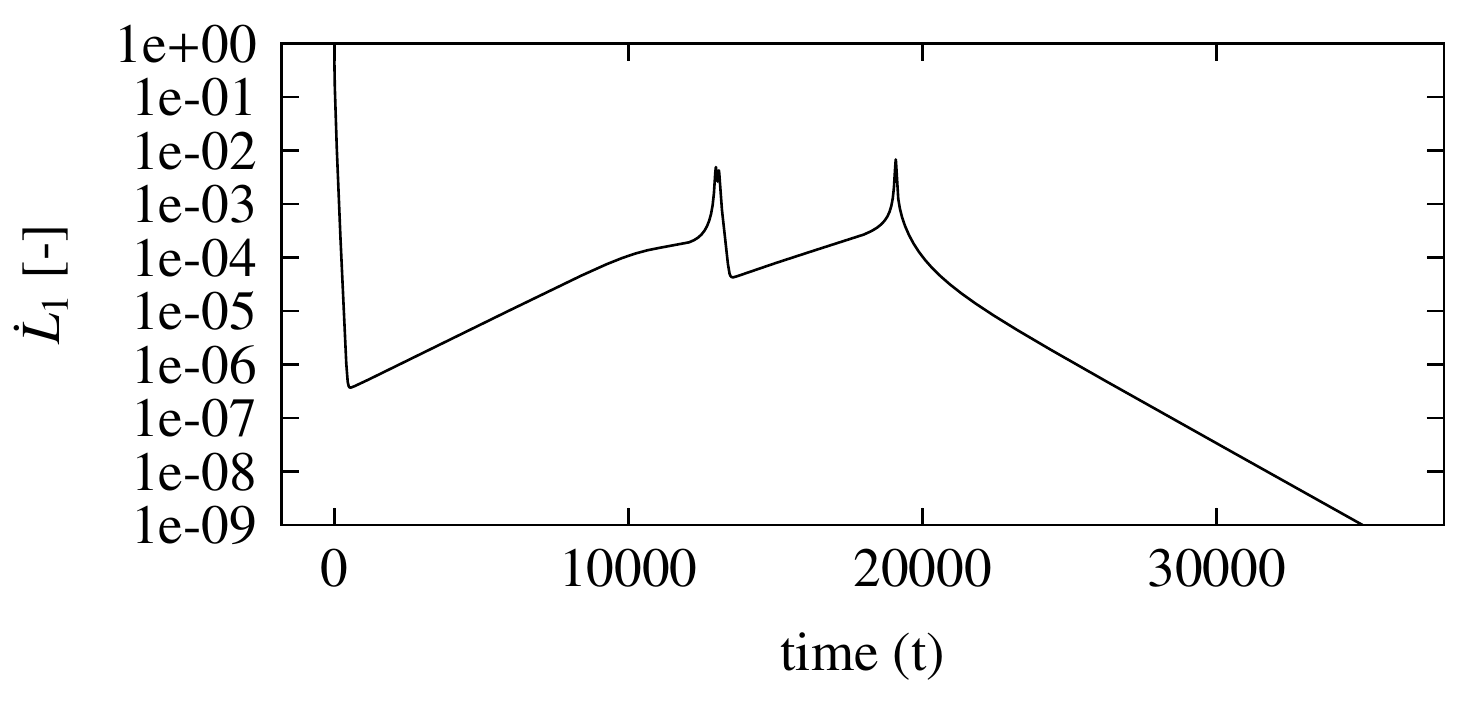}}\\[00pt]
		{(d)}
	\end{tabular}
	\caption{\label{fig:45 degrees preexist} 
		Pattern evolution from preexisting vertical stripes, subjected to PBC
		and a diagonal sinusoidal distribution of the control parameter, given
		by: $\varepsilon(\mathbf{x})=F\cos{[q_1 (x+y)]}$, where
		$q_1=0.125q_0$. For (a) $F=0.1$ and (b) $F=0.5$, the preexisting
		structure persists, and dominates the orientation effect of the
		gradient. (c) Upon adding a perturbation in the form of a uniform
		distribution ranging from $-10^{-2}$ to $10^{-2}$ to the initial
		condition, the preexisting structure collapses and is replaced by
		stripes parallel to the gradient. (d) The time evolution of
		$\dot{L}_1$ for the latter is shown and this simulation proceeded
		until $\dot{L}_1\leqslant 5 \times 10^{-9}$.}
	\end{figure}

	Fig.~\ref{fig:45 degrees preexist} presents a case of a preexisting
	structure of stripes along the $y$-direction and a sinusoidal profile
	of $\varepsilon$ along the diagonal of the domain, using PBC. Despite
	lacking the restrictive effect of boundary conditions, the gradient is
	dominated by the initial condition, and the preexisting structure
	persists. Upon adding a noise $\delta\psi \in (-10^{-2}, 10^{-2})$ to the
	initial condition, the preexisting structures is destabilized and
	replaced by a sinusoidal distribution of stripes parallel to the
	gradient.
	
	\subsection{Gaussian forcings, and pseudo-random initial conditions}\label{sec:gaussian}

\begin{figure}
		{\centering
		 \includegraphics[width=0.475\textwidth]{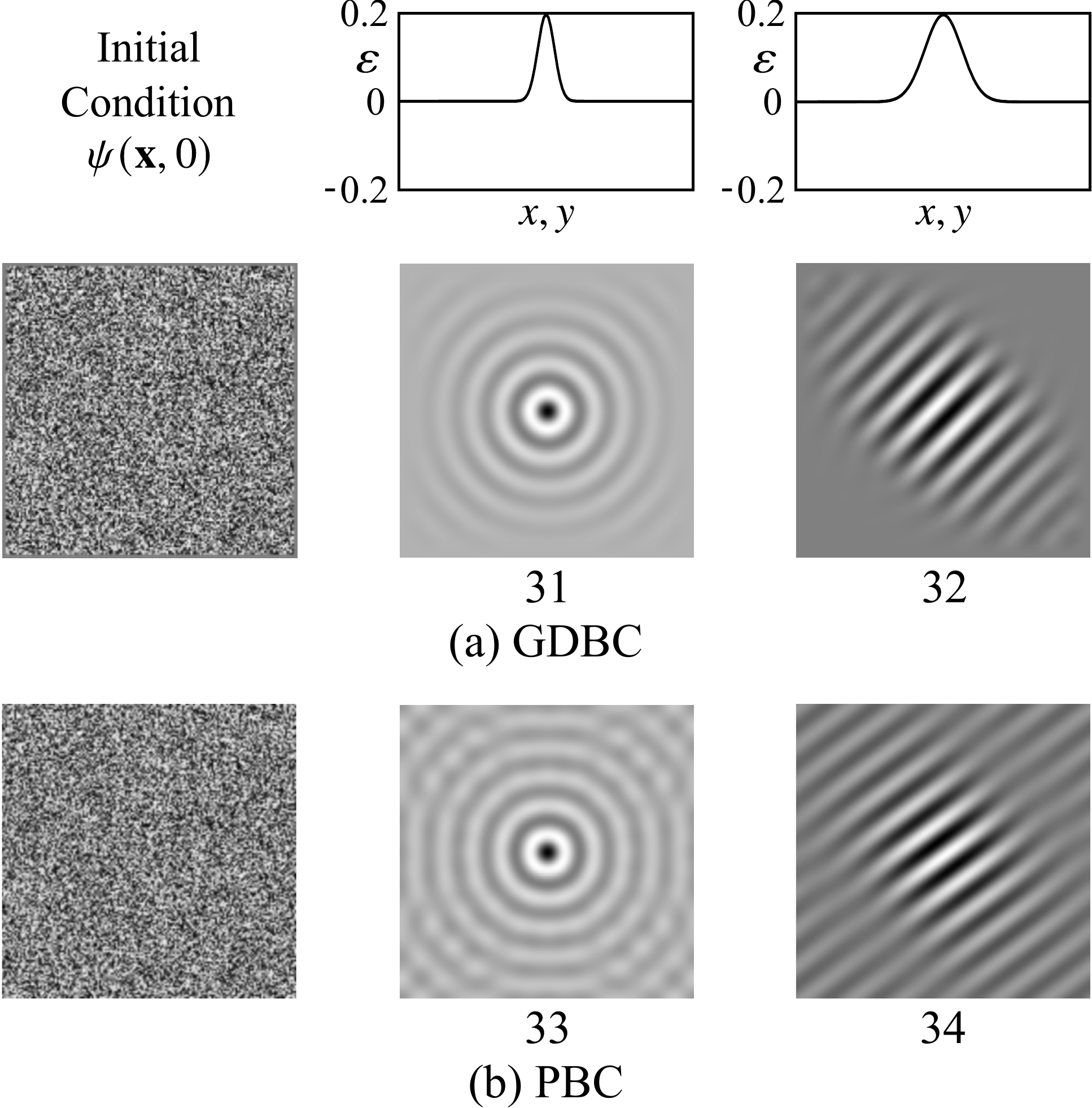}}
		\caption{\label{fig:gaussian}The results of four
		simulations with the SH3 equation, forced with a gaussian 
		distribution of $\varepsilon$. All simulations started
		from the same pseudo-random initial conditions, shown in the
		first column. Configurations 31 and 32 were run with GDBC, 
		while PBC were prescribed for configurations 33 and 33.}
	\end{figure}

	By imposing gaussian forcings we observe another bulk effect that
	competes with the gradient in orienting the stripes.
	Fig.~\ref{fig:gaussian} shows the steady state patterns obtained
	in four simulations, two of them run with a sharper circular gaussian
	distributions of the control parameter, centered at the middle of
	the domain and two, with a wider gaussian forcing. GDBC and PBC
	were considered for each configuration of $\varepsilon$. The four
	simulations started from the same pseudo-random initial condition
	adopted in all cases presented in Secs.~\ref{sec:random initial
		conditions} and~\ref{sec:comparison}. The adopted gaussian
	distribution is given by:
	\begin{equation}
	\label{eq:gaussian}
	\varepsilon\left(\bf{x}\right)=Ae^{-R((x-x_0)^2+(y-y_0)^2)} \;.
	\end{equation}
	
	In the first case, configurations 31 (GDBC) and 25 (PBC), we used a
	sharper distribution of $\varepsilon$ with parameters
	$R$ given in Tab.~\ref{tab:table2} ($R_1$ and $R_2$). For both, the
	resulting pattern takes the form of a target, with stripes
	presenting a wavevector parallel to the gradient, $\mathbf{q}
	\parallel \nabla\varepsilon$, a completely opposite situation with
	respect to several other cases run from pseudo-random initial
	conditions using forced with ramps or sinusoidal distributions of
	$\varepsilon$. The orientation effect of the gradient does not
	appear in this case, and orientation is dominated by a geometric
	bulk effect. Due to the disk-form of $\varepsilon$ in the
	two-dimensional domain, a target pattern is the one that fills the
	supercritical region while minimizing defects, which is a geometric
	compatibility effects. Otherwise, if stripes were to orient
	accordingly to the $\nabla\varepsilon$, the resulting pattern would
	contain a large amount of defects (dislocations) to accommodate
	such orientation, increasing significantly the energy of the
	configuration. Therefore, target patterns minimizes the Lyapunov
	potential, in spite of being penalized by the control parameter
	gradient. 
	
	A second case with a wider gaussian forcing is shown in
	Fig.~\ref{fig:gaussian}, configurations 32 (GDBC) and 33 (PBC), with
	parameters $R$ given in Tab.~\ref{tab:table2} ($R_1$ and $R_2$). This
	case corresponds to a forcing with a sharper distribution of Fourier
	modes, therefore a smaller range of modes persists in the steady state
	pattern. We observe that a pattern of stripes with wavevector aligned
	with the diagonal of the domain appears. The pattern extends for a
	longer distance in this diagonal direction, even invading the
	subcritical region, for both GDBC nad PBC. The effect is due to the
	fact that the hardest direction for modulation of the amplitude occurs
	in the direction of the wavevector, whereas the easiest direction is
	the perpendicular one. This property of periodic patterns results in
	amplitudes modulated in compliance with Newell-Whitehead-Segel
	equations~\cite{newell1969,Segel1969}.
	
	
	\section{
             Competition between the gradient, 
             boundary and bulk effects -- SH23 and SH35}\label{sec:SH35}

	In this section, we briefly explore two other forms of the SH equation with additional nonlinearities, which present different bifurcation diagrams. For instance, SH23 typically presents transition from the homogeneous state to hexagons close to $\varepsilon = 0$, and from hexagons to stripes as $\varepsilon$ increases, while the transition from the homogeneous state to stripes in SH35 is associated to a subcritical bifurcation at a negative $\varepsilon$. These transitions and alignment of resulting patterns may be affected by the presence of a control parameter gradient, leading to a more intricate interplay than the one discussed for SH3.
	As mentioned before,
	Eq.~\ref{SH} can be addressed as SH23 for $\gamma\eq0$,
	and $\zeta, \beta \neq 0$. Analogously, we refer to it as SH35 for
	$\zeta\eq0$, and $\gamma, \beta \neq 0$. The numerical scheme follows
	the semi-implicit approach described in 
	Appendix~\ref{appendix:numerical scheme} with minor modifications
	depending on which nonlinear terms are present.
	
\begin{figure}
		\centering {\includegraphics[width=0.475\textwidth]{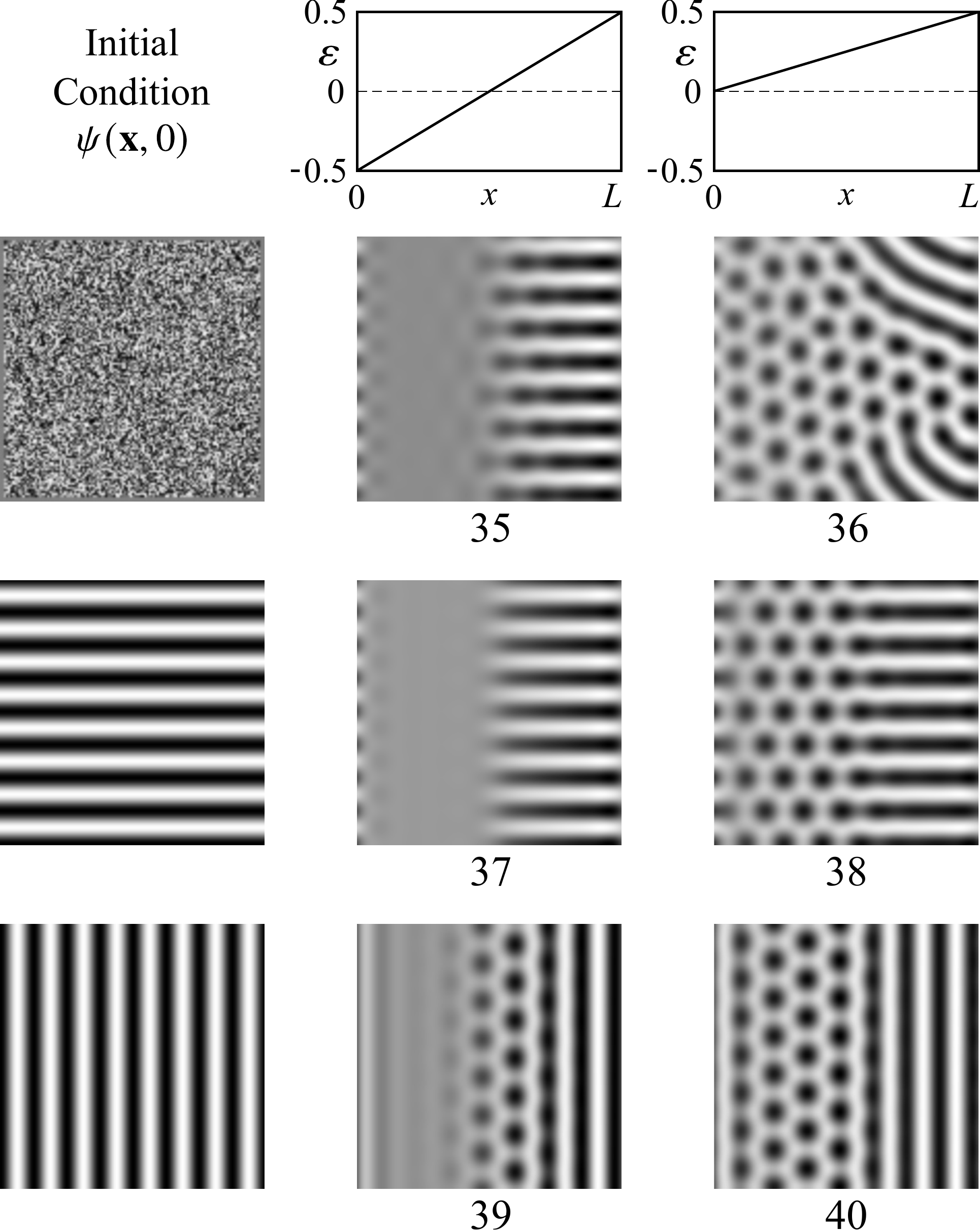}}
		{\caption{\label{fig:SH23} The results of six simulations with the
		SH23 equation, with PBC, forced with a spatial ramp of the control parameter
		$\varepsilon$ indicated in the first row. The parameters of the
		SH23 were: $\zeta = 0.65$, $\beta = 1$ and $\gamma = 1$.
		Initial conditions for configurations 37, 38, 39 and
		40 were perturbed so that the remaining stripes share the domain
		with hexagons, as expected for this range of parameters.}}
	\end{figure}

	Accordingly to the simulations presented in the previous sections
	and Refs.~\cite{pontes1994,C.I-1997,C.I-2002,pontes2008}, GDBC introduces
	additional restraints for the patterns, i.e, stripes anchoring
	perpendicularly to sidewalls. We adopt PBC to study interacting bulk
	effects for SH23 in the presence of a nonzero
	$\nabla\varepsilon$. For small positive $\varepsilon$,
	hexagonal patterns are the minimum energy state, which destabilizes when
	$\varepsilon$ is increased and stripe patterns become energetically favored.
	The coexistence of both structures with a nonuniform forcing was addressed
	by Hilali et al.~\cite{hilali1995}, where stripes formed with
	$\mathbf{q}\parallel\nabla\varepsilon$, contrary to our results for SH3.
	In order to clarify if such effect observed for SH23 was induced by initial
	conditions (an initial ramp in $\psi$, in their case), and assess 
	how $\nabla\varepsilon$ interferes in the hexagon to stripe transition,
	we perform simulations for SH23 using different initial conditions and
	ramps for $\varepsilon$. Numerical results are shown in Fig.~\ref{fig:SH23},
	using $\zeta = 0.65$ and $\beta = 1$, in which we observe the possibility
	of coexistence between hexagons and stripes.
	The nonuniform forcings considered were the following ramps:
	$-0.5\leqslant\varepsilon(\mathbf{x})\leqslant0.5$ and
	$0.0\leqslant\varepsilon(\mathbf{x})\leqslant0.5$. Three configurations
	for each forcing were considered, starting from pseudo-random initial 
	conditions and
	preexisting patterns (horizontal and vertical stripes, respectively).  
	
	Configurations 35 and 36 started from pseudo-random initial conditions,
	and configurations 37, 38, 39 and 40 had their preexisting condition
	perturbed with a pseudo-random noise ranging from $-10^{-6}$ to
	$10^{-6}$ with an uniform distribution. In the absence of perturbations,
	the initial condition is preserved, \emph{i.e}, hexagon patterns do
	not emerge. The results show that, in the presence of a
	$\nabla\varepsilon \neq 0$, the preexisting patterns are unstable, and
	we observe regions where stripes decay to $\psi = 0$ or evolve
	towards hexagons.
	
	First, we compare configurations 35, 37 and 39, where a ramp ranging
	from $\varepsilon = -0.5$ to $\varepsilon = 0.5$ was employed.
	Setting the initial condition as pseudo-random, we observe that
	stripes with $\mathbf{q}\perp\nabla\varepsilon$ appear in the region
	of positive $\varepsilon$, with a weakly formed hexagonal structure
	close to $\varepsilon = 0$. Configuration 37 show that when a
	preexisting structure of stripes with $\mathbf{q}\perp\nabla\varepsilon$
	is perturbed, stripes in the positive $\varepsilon$ region remain
	perfectly aligned to the gradient, and no transition to hexagons is
	observed. In configuration 39 we see that by perturbing an initial
	condition of stripes with $\mathbf{q}\parallel\nabla\varepsilon$,
	the remaining stripes did not reorient according to $\nabla\varepsilon$,
	and (opposite to configuration 37) a well formed column of hexagons appeared for
	regions of small $\varepsilon$. This is a consequence of the higher
	energy associated to stripes when $\mathbf{q}\parallel\nabla\varepsilon$,
	as compared in Fig. \ref{fig:Lya_curves}, so that for small $\varepsilon$ a
	stripe to hexagon transition is promoted. For configurations 36, 38 and
	40, where the ramp ranges from $\varepsilon = 0$ to $\varepsilon = 0.5$,
	we note that due to the smaller $\nabla\varepsilon$ this gradient
	has a weaker effect on inducing stripe pattern alignment. In
	configuration 36, we do not see alignment starting from a pseudo-random
	initial condition, while in configuration 38, even though the
	preexisting pattern was made of stripes with
	$\mathbf{q}\perp\nabla\varepsilon$, hexagons still emerged in the
	$0 < \varepsilon < 0.25$ region (opposite to configuration 37).
	
	Finally, we present numerical results for the SH35 equation in the
	presence of a control parameter gradient, using $\beta = 3$,
	and $\gamma = 1$. The $\psi = 0$ to stripe
	transition in SH35 is associated to a subcritical bifurcation,
	so that there is a jump of the amplitude in this transition.
	Due to the symmetry in the energy structure associated to SH35,
	the bifurcation parameter presents a coexistence value
	$\varepsilon_c = -27\beta^2/160\gamma$ for which both stripes and
	$\psi = 0$ states have approximately zero energy density
	\cite{sakaguchi1996,vitral2019role}. For $\varepsilon > \varepsilon_c$,
	stripes are energetically favored, while for $\varepsilon < \varepsilon_c$
	the equilibrium state is $\psi = 0$. A consequence of the subcritical
	bifurcation is that even in $\varepsilon > \varepsilon_c$ regions, stripes
	do not form from a pseudo-random initial condition for finite values of
	$\varepsilon_c$. Therefore, in the SH35 case we adopt a square
	pattern as initial condition, of the type
	$\psi  = A\textrm{cos}(q_0x)+B\textrm{cos}(q_0y)$, in order to evaluate
	the $\nabla\varepsilon$ effect on filtering the pattern. 
	
	For obtaining configurations 41 and 42 we use GDBC, while for
	configurations 43 and 44, we use PBC. For the chosen set of parameters
	$\varepsilon_c \approx -1.52$. With a ramp ranging from
	$\varepsilon = -3$ up to $\varepsilon = 0$, for both GDBC (away from
	the boundary) and PBC, the $y$ direction mode was filtered, and the
	resulting structures present stripes with
	$\mathbf{q}\perp\nabla\varepsilon$ for $\varepsilon > \varepsilon_c$,
	and $\psi = 0$ for $\varepsilon < \varepsilon_c$. By changing the ramp
	to $\varepsilon = -1.8$ up to $\varepsilon = 0$, we see in
	configuration 44 (PBC) that away from the boundaries the structure
	perfectly aligns according to $\nabla\varepsilon$. However, in
	configuration 42 we note that by decreasing the ramp inclination,
	orientation effects due to boundary conditions become stronger
	in comparison to $\nabla\varepsilon$, as they mostly dictate the
	resulting pattern. From our simulations using SH35 with PBC, we
	observed that patterns tend to orient more strongly according to
	$\nabla\varepsilon$ than in the case of SH3 or SH23, which is
	presumably associated to the subcritical nature of the bifurcation.
	
	\begin{figure}
		\begin{tabular}{c}
			{\includegraphics[width=0.475\textwidth]{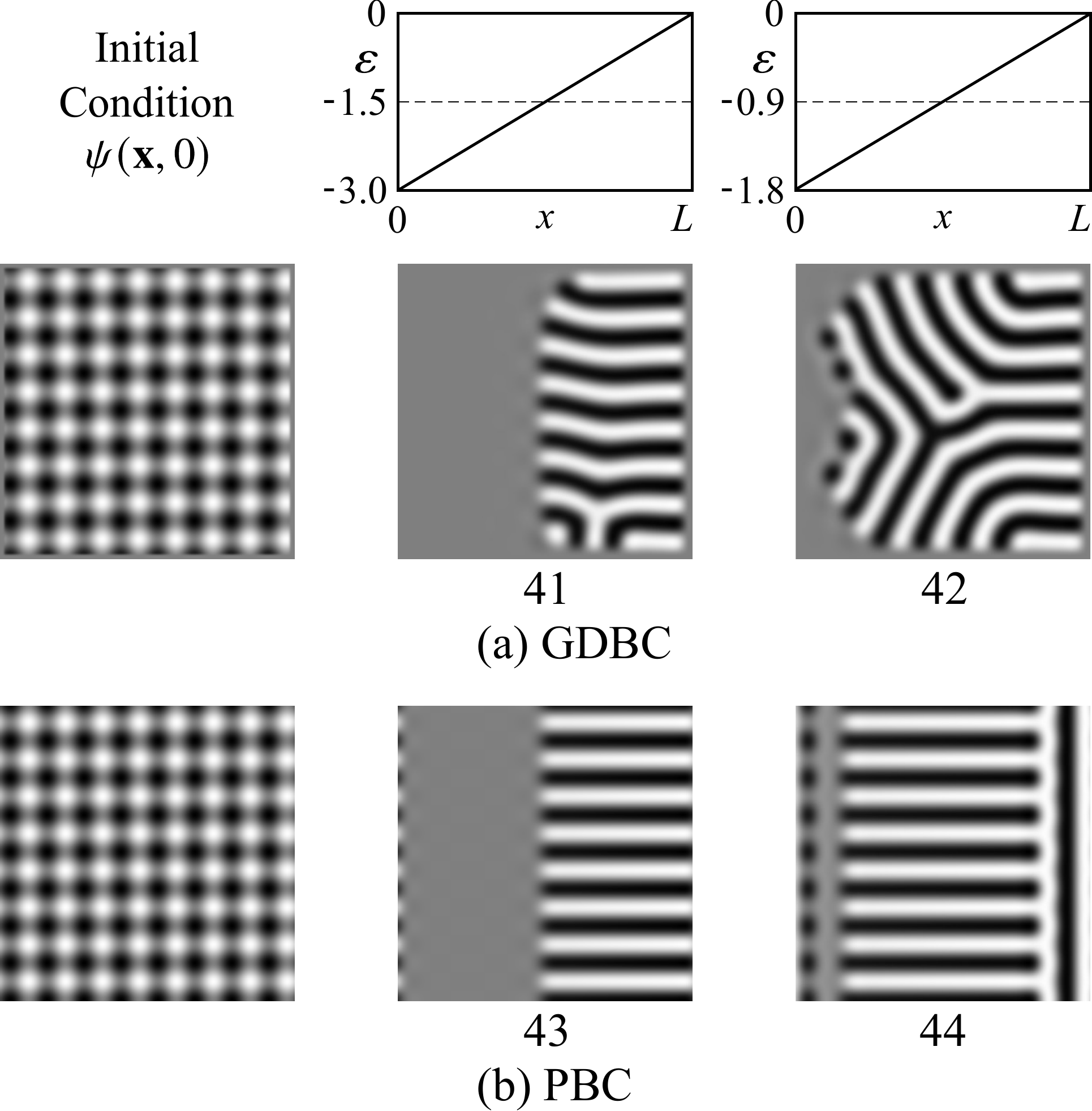}}
		\end{tabular}
		{\caption{\label{fig:SH35}
			The results of four
			simulations with the SH35 equation, forced with a spatial ramp of
			the control parameter $\varepsilon$. All simulations started
			from the same preexisting square pattern 
			($\psi  = A\textrm{cos}(q_0x)+B\textrm{cos}(q_0y)$), shown in the
			first column. The remaining columns present the steady state.
			First row: the prescribed profile of the control parameter
			$\varepsilon$. Second and third rows: results for GDBC, and PBC, 
			respectively. Results of the second row show
			that boundary effects dominate the orientation effect when GDBC
			are prescribed. In the third row, the absence of boundary
			effects allows the dominance of the orientation effect of the
			gradient.}}
	\end{figure}
	
	
	\section{Conclusions}\label{sec:conclusions}
	   
	This work addresses the stripe orientation effect due to the spatial gradient
	of the control parameter $\varepsilon$ (forcing) in the Swift-Hohenberg dynamics, 
	and how it fares against competing effects. In particular, we investigate
	an amplitude instability driven by the presence of a spatially inhomogeneous
	$\varepsilon$, which is not captured by classical studies of the stability
	of stripes such as the Busse balloon. We numerically show that stripes with
	wavevector $\mathbf{q}$ perpendicular to $\nabla\varepsilon$ are stable
	and correspond to a lower energy state than stripes with $\mathbf{q}$ parallel to
	$\nabla\varepsilon$, which are unstable near the bifurcation point.
	This is the fundamental result that explains the observed patterns for all
	the Swift-Hohenberg equations studied, independently of their nonlinearities
	(SH3, SH23, SH35). Not only stripes tend to show $\mathbf{q}$ perpendicular to
	$\nabla\varepsilon$ for both supercritical (SH3) and subcritical (SH35) bifurcations
	near the transition from the homogeneous state to stripes, but also
	$\nabla\varepsilon$ can modify a bifurcation diagram by favoring aligned stripes
	over hexagons (SH23). That is, if $\nabla\varepsilon$ is sufficiently high, a direct 
	transition from the homogeneous state to stripes becomes possible for a system
	which presents hexagons as an intermediary state between them. Further,
	we show that stripe reorientation by amplitude perturbation is possible if 
	the initial configuration is comprised of 
	stripes whose $\mathbf{q}$ is not perpendicular to $\nabla\varepsilon$, a result that agrees with analytic suggestions from our 
	weakly nonlinear analysis.
	
	However, our main numerical results show that
	the orientation effect of the control parameter gradient, despite existing, does not
	always prevail when facing competition with other bulk, boundary,
	geometric, and periodic effects due to computational domains.
	This competition leads to the emergence of a rich dynamics, as
	apparent in our results, which strongly depends on the magnitude of
	the forcing and initial conditions (preexisting patterns or
	pseudo-random). In this sense, the various forms of forcing and
	initial conditions addressed in this work, while extensive, do not
	cover the full range of possible cases. 
	What is clear is that the hierarchy of effects
	changes as a function of the $\nabla\varepsilon$ magnitude, so that
	the gradient effect may dominate over boundary conditions and bulk
	effects (e.g. by removing defects) if the magnitude becomes high. An 
	exception is when the forcing is a gaussian distribution, since the geometric compatibility
	effect leads to a target pattern of concentric rings, and any other configuration
	would imply in a high density of defects in order to accommodate the stripes.
	In summary, when $\nabla\varepsilon$ is small our results suggest
	geometry $>$ boundary conditions $>$ forcing level/defects $>$ $\nabla\varepsilon$ as the hierarchy
	on dictating the local orientation of stripes, whereas for high $\nabla\varepsilon$
	we observe geometry $>$ $\nabla\varepsilon$ $>$ boundary conditions $>$ forcing level/defects.

	This conditional hierarchy may prove helpful in pattern formation
	experiments seeking to overcome bias or defects introduced by boundary
	conditions and bulk effects, since it suggests that more uniformly
	oriented pattern may be achieved by tuning the control parameter field
	(e.g. texture control by increasing the temperature gradient). 
    This orientation effect is relevant 
	for many physical systems presenting periodic patterns, such
	as in developmental biology
	\cite{hiscock2015orientation,ruppert2020nonlinear}, smectic
	mesophases \cite{vitral2019role}, and localized sand patterns
	\cite{auzerais2016formation}, whose dynamics have been
	studied by Swift-Hohenberg type equations, but present mechanisms
	of stripe orientation that are not well understood. While we
	focus on a particular amplitude instability, the question
	remains on how gradients of the forcing would affect other
	instabilities, such as Eckhaus and zigzag. Future work could also
	address if the present observations are translated into three-dimensional
	phase-field models adopting Swift-Hohenberg type equations,
	and investigate how the velocity field in models presenting
	order parameter advection competes with $\nabla\varepsilon$ on the
	orientation of stripes.


	\section*{Acknowledgments}
	The authors thank FAPERJ (Research Support Foundation of the State
	of Rio de Janeiro) and CNPq (National Council for Scientific and
	Technological Development) for the financial support. Daniel Coelho
	acknowledges a fellowship from the Coordination for the Improvement
	of Higher Education Personnel-CAPES (Brazil). A FAPERJ Senior
	Researcher Fellowship is acknowledged by J. Pontes. The authors
	dedicate a special thanks to prof. D. Walgraef, from the Free 
	University of Brussels, who early pointed to the orientation effects
	of nonuniform forcings, on patterns of stripes. This endeavor is based 
	on his ideas.
	
	\appendix
	
	\section{Comparison between numerical and analytical results for the
	         amplitude -- additional results}\label{appendix:cross-sections}

\begin{figure}
	\begin{tabular}{c}
		{\includegraphics[width=0.475\textwidth]{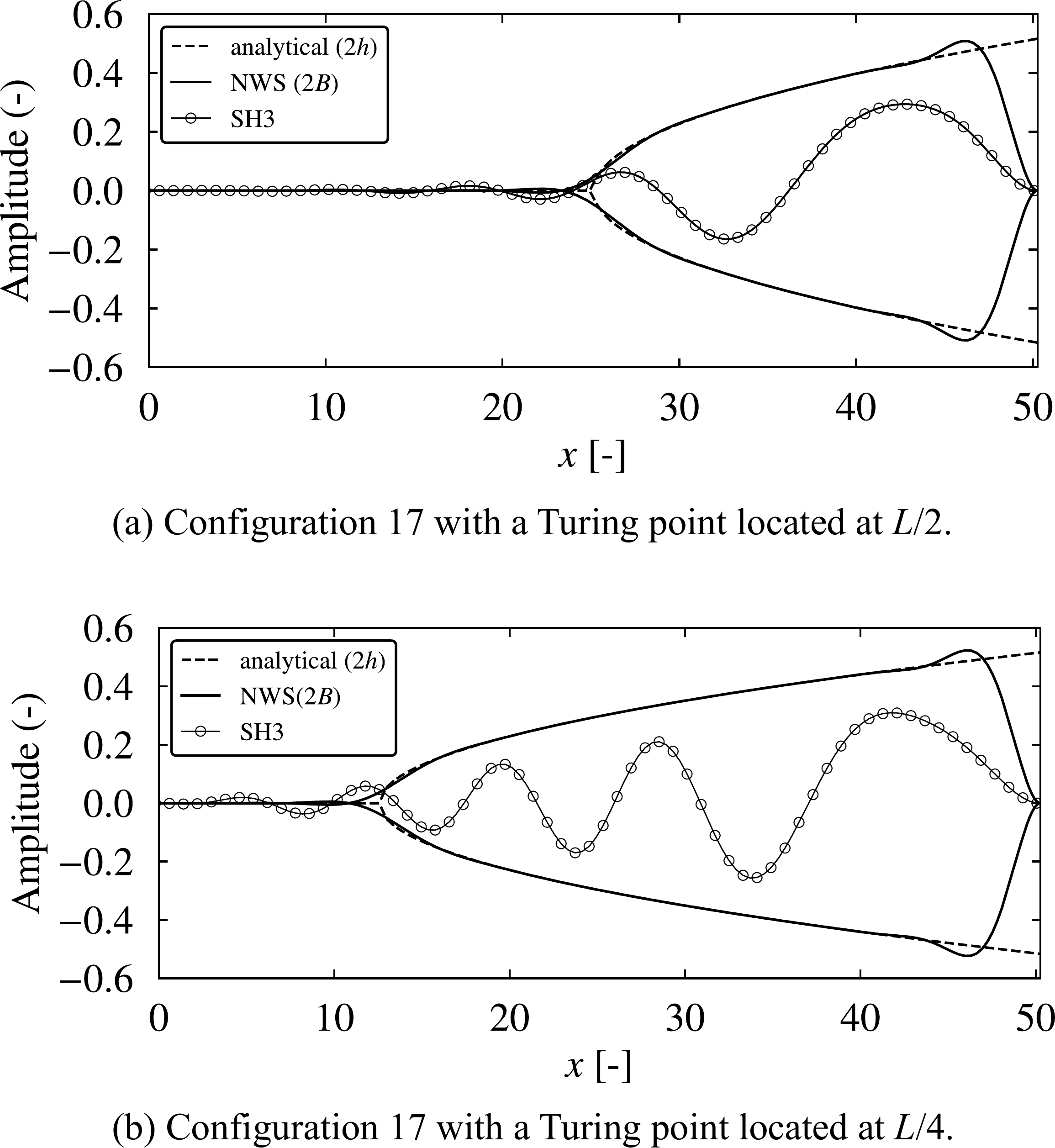}}
	\end{tabular}
    \caption{\label{fig:cross section subcritical}
        Cross section of configuration 17 of Fig.~\ref{fig:lyaresults2}, and 
        of the pattern obtained from the same configuration as initial
		condition, now run with a ramp of $\varepsilon$ where the bifurcation point
		is located at 1/4 of the domain length. The same maximum value
		of $\varepsilon$ prescribed for configuration 1 was adopted for the
		simulation shown in the second row of the present figure. Though
		associated to straight stripes aligned to the gradient, the envelopes
		fit well the crests of the patterns, which consist of bent stripes,
		with the angle between the local wavevector and the gradient
		continuously varying across the domain.}
	\end{figure}

	Additional numerical results for the amplitude of the striped patterns are shown, and we further evaluate how the analytic predictions fare in a system presenting a control parameter with spatial dependence.
	Figure~\ref{fig:cross section subcritical} shows in the first row
	the cross section of configuration 17 from Fig.~\ref{fig:lyaresults2}, whose forcing is a ramp in $x$.
	As in configuration 1 profile from Fig. \ref{fig:comparison2}, the
	evaluated envelopes refer to modes parallel to the gradient and not
	to bent stripes of the pattern, where the angle between the local
	wavevector and the gradient continuously varies across the domain.
	In spite of this fact, the envelopes qualitatively fit the peaks of the
	pattern, suggesting that the height of the envelope far from
	boundaries depends primarily on the local value of $\varepsilon$ and
	not on the direction of the wavevector.
	
	In the second row of Fig.~\ref{fig:cross section subcritical}, we move the bifurcation point from 1/2 to 1/4 of the
	domain length, in order to evaluate if a translation of the forcing
	would have any effect on the orientation of the pattern. We used
	the steady pattern of configuration 17 as the initial condition, for
	the configuration shown in the second row. In both cases, the cross
	section was taken at the middle height ($y$-direction) of the domain.
	We compare the $\psi$ profile with the estimated amplitude $B$,
	evaluated by the steady state numerical solution of the NWS Eq.\ref{eq:amp-b}, and also with asymptotic amplitude from Eq.~\ref{eq:h}.
	
	Note that the solution of the NWS Eq.~\ref{eq:amp-b} presents an
	estimate for the envelope on the subcritical region. The results show
	that a translation of the forcing ramp expands the pattern towards the
	new location of the bifurcation point, keeping the original bent form of the
	pattern in configuration 17. Therefore, bulk and boundary effects still
	win over gradient effects, and no reorientation is observed.
	
	\begin{figure}
		\centering
		{\includegraphics[width=0.475\textwidth]{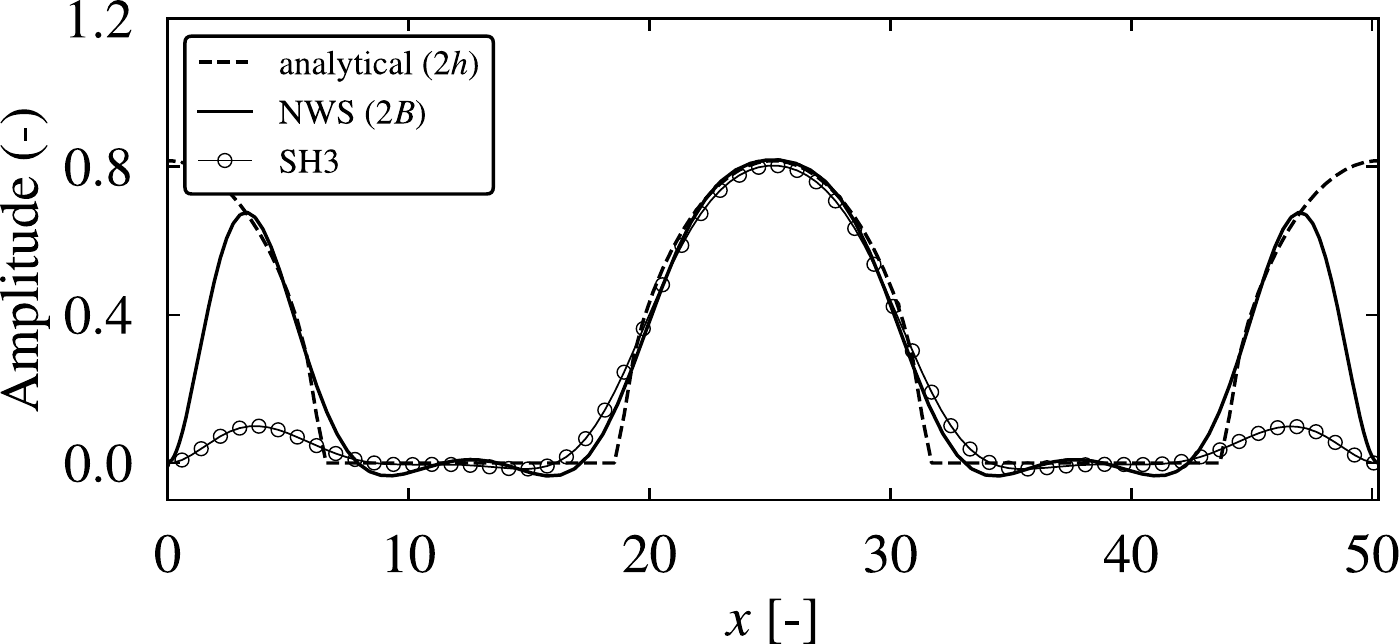}}
		{\caption{\label{fig:cross section sinusoidal}Cross
		section along the $x$-direction, of the pattern shown in
		configuration 22 of Fig.~\ref{fig:lyaresults3} (dotted line). The
		cross section was taken at the middle of the height ($y$-direction).
		To the profile obtained directly from the pattern we superposed the
		envelope of mode $B$, obtained by two methods: as the steady state
		solution of Eqs.~\ref{eq:amp-a2} and~\ref{eq:amp-b2}, adopting
		a sinusoidal distribution of $\varepsilon$ along the $x$ direction
		and as an estimation of the envelope profile, using Eq.~\ref{eq:h}.}}
	\end{figure}
	
	Fig.~\ref{fig:cross section sinusoidal} shows a cross section of the
	steady state pattern from configuration 30 of
	Fig.~\ref{fig:lyaresults3}, in the presence of a sinusoidal forcing. The cross section profile was taken
	at the middle of the height ($y$-direction) and is represented by a
	a line with small circles. To this profile we superposed the pattern envelope of
	mode $B$ estimated from two approaches: the first one consisted in
	the steady state solution of Eqs.~\ref{eq:amp-a2}
	and~\ref{eq:amp-b2}, for the amplitude $B$, using a sinusoidal
	distribution of $\varepsilon$. The so
	evaluated envelope is represented by a continuous line in
	Fig.~\ref{fig:cross section sinusoidal}. The second
	evaluation of the envelope was done by using Eq.~\ref{eq:h} also
	adopting a sinusoidal distribution of $\varepsilon$. The envelope
	is represented by a dashed line in
	Fig.~\ref{fig:cross section sinusoidal}. The two estimated envelopes
	fit the central part of the pattern cross section, and deviate at
	the boundaries, which are not taken into account in the derivation
	of those two curves.
	
	
	\section{\label{appendix:numerical scheme}Finite difference scheme 
	         for solving the bidimensional Swift-Hohenberg equation}
		The cubic, quadratic-cubic and cubic-quintic (SH3, SH23, SH35,
	respectively) equations were solved by the finite difference method in
	uniform and structured meshes, with a semi-implicit Crank-Nicolson
	scheme. The scheme is unconditionally stable, features truly second
	order representation of time and space derivatives, and strict
	representation of the associated Lyapunov functional, which
	monotonically decays. Details of the scheme are given by Christov and
	Pontes (2002)~\cite{C.I-2002} and by Coelho \emph{et
		al.}~(2020)\cite{coelho2020}. The scheme is summarized below, for the
	sake of completeness. 
	
	\subsection*{Target scheme}\label{sec:the target scheme}
	In order to construct a
	Crank-Nicolson second order in time numerical scheme, we adopt the
	following discrete representation of Eq.~\ref{SH}:
	\begin{eqnarray}
		&&
		\nonumber
		\frac{\psi^{n+1}-\psi^{n}}{\Delta t}
		\eq
		\left[
		\varepsilon(\mathbf{x})-\alpha q_0^4
		-
		2\alpha q_0^2\partial_x^2
		-
		2\alpha q_0^2\partial_y^2
		-
		\right.
		\\
		\nonumber
		&&
		\left.
		\alpha\partial_x^4
		-
		2\alpha\partial_x^2\partial_y^2
		-
		\alpha\partial_y^4
		+
		\zeta\frac{\left(\psi^{n+1}\right)
			+
			\left(\psi^n\right)}{2}
		+
		\beta\frac{\left(\psi^{n+1}\right)^2
			+
			\left(\psi^n\right)^2}{2}
		\right.
		\\
		&&
		\left.
		-
		\gamma\frac{\left(\psi^{n+1}\right)^4
			+
			\left(\psi^n\right)^4}{2}
		\right]
		\left(
		\frac{\psi^{n+1}+\psi^{n}}{2}
		\right).
		\label{operatorlambda}
	\end{eqnarray}
	The superscript $(n+1)$ refers to variables evaluated at the next
	time step, and $n$, to the ones evaluated at the current one.
	The RHS of Eq.~\ref{operatorlambda} consists of a nonlinear operator
	(in braces) actuating on the order parameter $\psi$, evaluated as
	the mean value at the middle of the time steps $(n)$ and $(n+1)$.
	The terms in Eq.~\ref{operatorlambda} are grouped in the target
	scheme, as follows:
	\begin{equation}
		\label{eq:target scheme}
		\frac{\psi^{n+1}-\psi^{n}}{\Delta t}
		\eq
		\left(2\Lambda_{x}^{n+1/2}+2\Lambda_{y}^{n+1/2}\right)
		\frac{\psi^{n+1}+\psi^{n}}{2}+f^{n+1/2}.
	\end{equation}
	
	For the SH23 and SH3
	equations ($\gamma=0$), the operators $\Lambda_{x}^{n+1/2}$,
	$\Lambda_{y}^{n+1/2}$ and $f^{n+1/2}$ are defined as:
	\begin{eqnarray*}
		\Lambda_{x}^{n+1/2}
		&=&
		\frac{1}{2}
		\left[
		-\alpha
		\left(\partial_x^4+\frac{q_0^4}{2}\right)
		-
		\beta
		\frac{\left(\psi^{n+1}\right)^2+\left(\psi^n\right)^2}{2}
		\right];
		\\
		\Lambda_{y}^{n+1/2}
		&=&
		\frac{1}{2}
		\left[
		-\alpha
		\left(\partial_y^4+\frac{q_0^4}{2}\right)
		-
		\beta
		\frac{\left(\psi^{n+1}\right)^2+\left(\psi^n\right)^2}{2}
		\right];
		\\
		f^{n+1/2}
		&=&
		\frac{1}{2}
		\left[
		\varepsilon(\mathbf{x})
		-\alpha
		\left(
		2q_0^2\partial_x^2+2q_0^2\partial_y^2+
		\right.
		\right.
		\\
		&&
		\left.
		\left.
		2\partial_x^2\partial_y^2
		\right)
		+
		\zeta
		\frac{\left(\psi^{n+1}\right)+\left(\psi^n\right)}{2}
		\right]
		\left(\psi^{n+1}+\psi^{n}\right),
		\label{eq:operators SH3}
	\end{eqnarray*}
	and for the SH35 equation ($\zeta=0$):
	\begin{eqnarray*}
		\Lambda_{x}^{n+1/2} &=&
		\frac{1}{2}
		\left[
		-\alpha
		\left(\partial_x^4+\frac{q_0^4}{2}\right)
		-
		\gamma
		\frac{\left(\psi^{n+1}\right)^4+\left(\psi^n\right)^4}{2}
		\right];
		\nonumber
		\\
		\Lambda_{y}^{n+1/2} &=&
		\frac{1}{2}
		\left[
		-\alpha
		\left(\partial_y^4+\frac{q_0^4}{2}\right)-
		\gamma
		\frac{\left(\psi^{n+1}\right)^4+\left(\psi^n\right)^4}{2}
		\right];
		\nonumber
		\\
		f^{n+1/2} &=&
		\frac{1}{2}
		\left[
		\varepsilon(\mathbf{x})
		-
		\alpha
		\left(
		2q_0^2\partial_x^2+	2q_0^2\partial_y^2
		+
		2\partial_x^2\partial_y^2
		\right)
		\right.
		\nonumber
		\\
		&&
		\left.
		+
		\beta
		\frac{\left(\psi^{n+1}\right)^2+\left(\psi^n\right)^2}{2}
		\right]
		\left(\psi^{n+1}+\psi^{n}\right).
		\label{eq:operators SH35}
	\end{eqnarray*}
	Terms are assigned to $\Lambda_x^{n+1/2}$ and $\Lambda_y^{n+1/2}$ in
	order to construct negative definite operators that assure the
	unconditional stability of the scheme. The scheme is nonlinear and
	requires internal iterations at each time step. Terms with superscript
	$(n+1)$ in
	the function $f^{n+1/2}$ and in both operators, are evaluated at the
	current internal iteration, whereas the remaining ones ($\psi^{n+1}$ in
	Eq.~\ref{eq:target scheme}) are the truly implicit ones, evaluated at
	the next internal iteration.
	\subsection*{The splitting}\label{sec:the splitting scheme}
	The original scheme is replaced by two following equations, using the
	Stabilizing Correction scheme. The two equations are equivalent to the
	target scheme up to a second order error in the evaluation of the time
	derivative, and does not change the steady state solution. In addition,
	the two-equations scheme minimize memory requirements and truncations
	errors~\cite{C.I-2002,coelho2020}:
	\begin{eqnarray*}
		\frac{\widetilde{\psi}-\psi^{n}}{\Delta t}
		&=&
		\Lambda_{x}^{n+1/2}\widetilde{\psi}+\Lambda_{y}^{n+1/2}\psi^{n}
		+
		\\
		&&
		f^{n+1/2}+(\Lambda_{x}^{n+1/2}+\Lambda_{y}^{n+1/2})\psi^{n};
		\label{SS1}
		\\
		\frac{\psi^{n+1}-\widetilde{\psi}}{\Delta t}
		&=&
		\Lambda_{y}^{n+1/2}(\psi^{n+1}-\psi^{n}),
		\label{SS2}
	\end{eqnarray*}
	where $\widetilde\psi$ is an intermediary estimation of $\psi$ at the
	new time step. The second equation provides a correction and thus we
	obtain $\psi$ for the new time. Variables in the RHS of both equations
	are evaluated as the mean value between the time step $n$ and the
	estimation for the new one, at the previous internal iteration. All
	spatial derivatives are represented by second order central difference
	formul{\ae}.
	\subsection*{Steady state criteria}\label{sec:L1}
	We  follow the structure evolution by assessing the
	rate of change in time of the pattern during the simulation
	by monitoring $\dot{L}_1$, the relative norm rate of change defined
	as:
	\begin{equation}
		\dot L_1\;=\frac{1}{\Delta t}\;
		\left(
		\frac{
			\sum\limits_{i}^{n_x}\sum\limits_{j}^{n_y}
			\mid\psi_{i,j}^{n+1}-\psi_{i,j}^{n}\mid
		}
		{
			\sum\limits_{i}^{n_x}\sum\limits_{j}^{n_y}
			\mid\psi_{i,j}^{n+1}\mid}
		\right).
		\label{eq:L1}
	\end{equation}
	It roughly corresponds to the ratio between the spatial average of the
	modulus of time derivative ${\partial \psi}/{\partial t}$ and the spatial
	average of the modulus of the function itself. Furthermore, it is sensitive
	not only to the growth of the amplitude, but also to the pattern phase
	dynamics. The simulations proceeded until $\dot{L}_1\leqslant 5 \times 10^{-7}$,
	which is our criterion for reaching the steady state.
	\begin{table}
	\label{tab:table1}
		\caption{Final times required to reach the steady state criteria
			     for each configuration.}
		\begin{tabular}{@{}clllcll@{}} 
			\\\hlineB{2.5}
			\textbf{Config.} && \textbf{Steady State} && \textbf{Config.} 
			&& \textbf{Steady State}
			\\
			\hline
			01 && $t=43932$  && 23 && $t=31986$
			\\
			02 && $t=7404$  && 24 && $t=66862$
			\\
			03 && $t=297$  && 25 && $t=15911$
			\\
			04 && $t=82$  && 26 && $t=52611$
			\\
			05 && $t=21441$  && 27 && $t=3861$
			\\
			06 && $t=1968$  && 28 && $t=27641$
			\\
			07 && $t=4684$  && 29 && $t=4970$
			\\
			08 && $t=1459$  && 30  && $t=3980$
			\\
			09 && $t=119095$  && 31 && $t=5780$
			\\
			10 && $t=53560$    && 32 && $t=319880$
			\\
			11 && $t=2470$&& 33 && $t=8325$
			\\
			12 && $t=2623$  && 34 && $t=84100$
			\\
			13 && $t=160105$  && 35  && $t=18152$
			\\
			14 && $t=9371$  && 36  && $t=7365.5$
			\\
			15 && $t=9582$    && 37 && $t=296$
			\\
			16 && $t=9755$  && 38 && $t=24004$
			\\
			17 && $t=30970$  && 39 && $t=890$
			\\
			18 && $t=11885$  && 40 && $t=2338.5$
			\\
			19 && $t=12073.5$  && 41 && $t=5000$
			\\
			20 && $t=11606$  && 42 && $t=4500$
			\\
			21 && $t=7858$  && 43 && $t=106.4$
			\\
			22 && $t=4455$  && 44 && $t=131.6$
			\\
			\hlineB{2.5}
		\end{tabular}
	\end{table}
	
	\subsection*{The discrete implementation of the Lyapunov 
	functional}\label{sec:the discrete Lyapunov functional}

	The Lyapunov
	functional associated to the SH equation is implemented through the
	discrete formula derived by Christov \& Pontes~(2001) \cite{C.I-2002}
	for the SH3, and extended for the SH35 by Coelho \emph{et
		al.}~(2020) \cite{coelho2020}. The formula presents a ${\cal
		O}\left(\Delta t^2+\Delta x^2+ \Delta y^2\right)$ approximation of the
	functional given by Eq.~\ref{eq:dec_LF}:
	\begin{widetext}
		\begin{eqnarray}
			\frac{{\cal F}^{n+1} - {\cal F}^n}{\Delta t}
			\eq&&
			-\sum\limits_{i=1}^{n_x}\sum\limits_{j=1}^{n_y}
			\left({\psi^{n+1}_{i,j}- \psi^n_{i,j}\over \Delta t}\right)^2\;;\!\!
			\nonumber
			\\
			{\cal F}^n
			\eq&&
			\sum\limits_{i=1}^{n_x}\sum\limits_{j=1}^{n_y}
			\left[
			-{\epsilon\over 2}\left(\psi^{n}_{i,j}\right)^2
			-
			{\zeta\over 3}
			\left(\psi^{n}_{i,j}\right)^3
			-
			{\beta\over 4}
			\left(\psi^{n}_{i,j}\right)^4
			+
			{\gamma\over 6}
			\left(\psi^{n}_{i,j}\right)^6
			+
			{\alpha q_0^4\over 2}
			\left(\psi^{n}_{i,j}\right)^2
			\right]
			\nonumber
			\\
			&&
			-\frac{\alpha q_0^2}{2} 
			\sum\limits_{i=1}^{n_x}\sum\limits_{j=1}^{n_y}
			\left[
			{\psi^{n}_{i+1,j}-\psi^{n}_{i,j}\over \Delta x}
			\right]^2
			\!\!\!+\!
			\left[
			{\psi^{n}_{i,j}-\psi^{n}_{i-1,j}\over \Delta x}
			\right]^2
			\!\!\!+\!
			\left[
			{\psi^{n}_{i,j+1}-\psi^{n}_{i,j}\over \Delta y}
			\right]^2
			\!\!\!+\!
			\left[
			{\psi^{n}_{i,j}-\psi^{n}_{i,j-1}\over \Delta y}
			\right]^2
			\nonumber
			\\
			&&
			+
			\frac{\alpha}{2}
			\sum\limits_{i=1}^{n_x}\sum\limits_{j=1}^{n_y}
			\left[
			{\psi^{n}_{i+1,j}-2\psi^{n}_{i,j}+\psi^{n}_{i-1,j}
				\over \Delta x^2}
			+
			{\psi^{n}_{i,j+1}-2\psi^{n}_{i,j}+\psi^{n}_{i,j-1}
				\over \Delta y^2}
			\right]^2.
			\label{eq:Lyapunovfunctional}
		\end{eqnarray}
	\end{widetext}
	
	The monotonic decay of the
	finite difference version is enforced, provided that the internal
	iterations converge~\cite{C.I-2002}.
	
	\subsection*{Parameters adopted in the simulations}\label{sec:parameters}
	
	The simulations shared common parameters regarding the wavenumber,
	domain sizes and time steps. Those are presented in the following table
	along with the parameters used in particular simulations, such as the 
	gaussians distributions parameters.
	
	\begin{table}
		\label{tab:table2}
		\caption{Parameters adopted in the numerical study throughout
			     this work.}
		\begin{tabular}{@{}cccccll@{}}
			\\\hlineB{2.5}
			\textbf{Parameter} && \textbf{Formul\ae} && \textbf{Value} 
			&& \textbf{Description}
			\\
			\hline
			$q_0$ && -   && 1.0        && Critical wavenumber
			\\
			$\lambda_0$  && $2\pi/q_0$ && $2\pi$ && Critical wavelength
			\\
			$w_x$, $w_y$ &&  - && 8   && Wavelengths
			\\
			&&&&&&per domain length
			\\
			$g_r$  && - && 16 && Grid resolution
			\\
			$n_x$, $n_y$ &&  $w_x\times g_{r}$,&& 128
			&& Nodes per mesh side
			\\
			&&$w_y\times g_{r}$ && 
			\\
			$N$    &&  $n_x\times n_y$ && $128\times 128$
			&& Total mesh nodes
			\\
			$L_x$, $L_y$ &&  $w_x \lambda_0$, $w_y \lambda_0$ &&
			$\approx 50.2655$  &&Domain length ($L$)
			\\
			$\Delta x$,$\Delta y$ &&   $L/(n-2)$ && $\approx 0.3989$ &&
			Space step (GDBC)
			\\
			$\Delta x$,$\Delta y$ &&   $L/n$ && $\approx 0.3927$     &&
			Space step (PBC)
			\\
			$\Delta t$  &&  - && 0.5 && Time step (SH3)
			\\
			$\Delta t$  &&  - && 0.1 && Time step (SH23/SH35)
			\\
			$A$  && - && 0.2 && Gaussian maximum
			\\
			&&&&&& value (peak)
			\\
			$R_1$  && $n^{-1}_x$ && - && Configs. 04 and 09
			\\
			$R_2$  && $0.2n^{-1}_x$ && - && Configs. 05 and 10
			\\
			$x_0$,$y_0$  && $L_x/2$,$L_y/2$ && - && Gaussian center
			\\
			\hlineB{2.5}
		\end{tabular}
	\end{table}
	
	Space and time steps were chosen ~following \cite{coelho2020} for a good
	compromise between accuracy and computational cost (performance). The grid
	resolution represents the number of nodes per critical wavelength.
	
	\section{Pseudo-spectral schemes for solving the one dimensional
	         Newell-Whitehead-Segel equation}\label{appendix:numerical scheme2}
		In the weakly nonlinear analysis section, a pair of coupled
	Newell-Whitehead-Segel (NWS) equations (Eqs.~\ref{eq:amp-a} and \ref{eq:amp-b}) is
	derived for the modes $A(\mathbf{x},t)$ and $B(\mathbf{x},t)$ via the
	multiple scale formalism. In order to compare the amplitude envelopes
	from SH simulations and the ones described by those amplitude
	equations, we develop numerical solutions for the NWS. The
	one-dimensional ($x$-direction) NWS equation has the form,
	\begin{equation}
		\label{eq:NWSapp}
		\partial_t u=\varepsilon(\mathbf{x}) u
		-\alpha\partial_x^4 u
		+3\beta u^3\;,
	\end{equation}
	\noindent
	where $\alpha=1$, $\beta=-1$, $t\geqslant0$ and $u \equiv u(x,t)$ is a
	real function described in the regular domain $\Omega:\{x\in [0,L_x]\}$
	with periodic boundary conditions (PBC) and generalized Dirichlet
	boundary conditions (GDBC). Since an analytical study is not trivial
	for this equation, we develop a numerical study using a semi-implicit
	pseudo-spectral method with first-order accuracy in time. The Fourier
	approach was adopted for the configurations with periodic boundary
	conditions (PBC). The Chebyshev approach was adopted for the
	configurations subjected to generalized Dirichlet boundary conditions
	(GDBC), for dealing with the imposed boundary conditions. Both
	approaches are briefly discussed in the following sections.

	\subsection*{Fourier pseudo-spectral scheme}\label{sec:fourier}
	
	The eigenfunctions of the fourth-order differential operator over the
	domain with periodic boundary conditions (PBC) are the Fourier modes
	$e^{ik\cdot x}$ ($k\in\mathbb{Z}^N$). Since $\partial_x^4 e^{ik\cdot
		x}=|k|^4 e^{ik\cdot x}$, ~equation \ref{eq:NWSapp} can be written as
	\begin{equation}
		\partial_t \hat{u}_k
		=
		-\alpha |k|^4 \hat{u}_k+\hat{f}_k\;,
	\end{equation}
	\noindent
	where $\hat{u}_k$ is the Fourier coefficient associated with the mode
	$k$, and $\hat{f}_k$ is the Fourier transform of the nonlinear terms. 
	The fourth order 
	derivative term is treated implicitly since it
	has a numerical stabilizing property (denoted by index $n+1$). The
	control parameter term is explicit since it is destabilizing in
	the scheme (denoted by index $n$).  The nonlinear terms are computed in
	real space in order to avoid computing Fourier mode convolutions
	(higher computational effort), and therefore are treated explicitly.
	Since we are interested in the steady state solution, a semi-implicit
	first order accurate in time scheme was employed, and can be expressed
	as follows:
	\begin{align}
		\hat{u}_k^{n+1}
		&\eq
		\mu_k\left(\hat{u}_k^{n}+\Delta t\hat{f}_k^n\right)\;,
	\end{align}
	
	where $\mu_k=\left(1+\alpha \Delta t k^4\right)^{-1}$ and
	$\hat{f}_k^n=\mathcal{F}\{f(u^n)\}_k$ is the Fourier transform of the
	nonlinear and variable coefficient terms
	$f(u^n)=\varepsilon(\mathbf{x})u^n+3\beta(u^n)^3$. This transformation
	is performed via a fast Fourier transform (FFT)-based code without
	dealiasing, using Octave FFT library.
	
	\subsection*{Chebyshev pseudo-spectral scheme}\label{sec:cheb}
	
	Equation \ref{eq:NWSapp} is solved using Chebyshev spectral collocation
	~method \cite{boyd2001chebyshev}. Chebyshev polynomials of degree $n$
	have $n$ zeros in  the interval $\xi\in[-1,1]$ that should be mapped to
	the physical domain $x\in[0,L_x]$.  For that purpose, a simple mapping
	is chosen: $x=0.5(\xi+1)L_x$. The numerical scheme can be expressed as
	\begin{align}
		\left(I+\alpha\Delta t D_{x}^{4}\right)u^{n+1}
		&\eq
		u^{n}+\Delta t f^n\;,
	\end{align}
	\noindent
	where $f^n=\varepsilon(\mathbf{x})u^n+3\beta(u^n)^3$ and $D_{x}^{4}$ is
	the Chebyshev collocation for the fourth-order differential operator on
	the mapped domain. This system of linear equations is solved subjected
	to the boundary conditions for the amplitude $B$. For such purpose we
	adopted $B=\partial B/\partial x=0$, for the boundary points $x=0$ and
	$x=L_x$. These conditions are consistent with the GDBC used in the SH
	equation simulations, for the order parameter $\psi$.
    \bibliography{main}

\begin{thebibliography}{47}%
\makeatletter
\providecommand \@ifxundefined [1]{%
 \@ifx{#1\undefined}
}%
\providecommand \@ifnum [1]{%
 \ifnum #1\expandafter \@firstoftwo
 \else \expandafter \@secondoftwo
 \fi
}%
\providecommand \@ifx [1]{%
 \ifx #1\expandafter \@firstoftwo
 \else \expandafter \@secondoftwo
 \fi
}%
\providecommand \natexlab [1]{#1}%
\providecommand \enquote  [1]{``#1''}%
\providecommand \bibnamefont  [1]{#1}%
\providecommand \bibfnamefont [1]{#1}%
\providecommand \citenamefont [1]{#1}%
\providecommand \href@noop [0]{\@secondoftwo}%
\providecommand \href [0]{\begingroup \@sanitize@url \@href}%
\providecommand \@href[1]{\@@startlink{#1}\@@href}%
\providecommand \@@href[1]{\endgroup#1\@@endlink}%
\providecommand \@sanitize@url [0]{\catcode `\\12\catcode `\$12\catcode
  `\&12\catcode `\#12\catcode `\^12\catcode `\_12\catcode `\%12\relax}%
\providecommand \@@startlink[1]{}%
\providecommand \@@endlink[0]{}%
\providecommand \url  [0]{\begingroup\@sanitize@url \@url }%
\providecommand \@url [1]{\endgroup\@href {#1}{\urlprefix }}%
\providecommand \urlprefix  [0]{URL }%
\providecommand \Eprint [0]{\href }%
\providecommand \doibase [0]{http://dx.doi.org/}%
\providecommand \selectlanguage [0]{\@gobble}%
\providecommand \bibinfo  [0]{\@secondoftwo}%
\providecommand \bibfield  [0]{\@secondoftwo}%
\providecommand \translation [1]{[#1]}%
\providecommand \BibitemOpen [0]{}%
\providecommand \bibitemStop [0]{}%
\providecommand \bibitemNoStop [0]{.\EOS\space}%
\providecommand \EOS [0]{\spacefactor3000\relax}%
\providecommand \BibitemShut  [1]{\csname bibitem#1\endcsname}%
\let\auto@bib@innerbib\@empty
\bibitem [{\citenamefont {Swift}\ and\ \citenamefont
  {Hohenberg}(1977)}]{Swift-1977}%
  \BibitemOpen
  \bibfield  {author} {\bibinfo {author} {\bibfnamefont {J.}~\bibnamefont
  {Swift}}\ and\ \bibinfo {author} {\bibfnamefont {P.~C.}\ \bibnamefont
  {Hohenberg}},\ }\href {\doibase 10.1103/PhysRevA.15.319} {\bibfield
  {journal} {\bibinfo  {journal} {Physical Review A}\ }\textbf {\bibinfo
  {volume} {15}} (\bibinfo {year} {1977}),\
  10.1103/PhysRevA.15.319}\BibitemShut {NoStop}%
\bibitem [{\citenamefont {Brazovskii}\ and\ \citenamefont
  {Dmitriev}(1975)}]{brazovskii1975phase}%
  \BibitemOpen
  \bibfield  {author} {\bibinfo {author} {\bibfnamefont {S.}~\bibnamefont
  {Brazovskii}}\ and\ \bibinfo {author} {\bibfnamefont {S.}~\bibnamefont
  {Dmitriev}},\ }\href@noop {} {\bibfield  {journal} {\bibinfo  {journal} {Zh.
  Eksp. Teor. Fiz}\ }\textbf {\bibinfo {volume} {69}},\ \bibinfo {pages} {979}
  (\bibinfo {year} {1975})}\BibitemShut {NoStop}%
\bibitem [{\citenamefont {Manneville}(1983)}]{manneville1983two}%
  \BibitemOpen
  \bibfield  {author} {\bibinfo {author} {\bibfnamefont {P.}~\bibnamefont
  {Manneville}},\ }\href@noop {} {\bibfield  {journal} {\bibinfo  {journal}
  {Journal de Physique}\ }\textbf {\bibinfo {volume} {44}},\ \bibinfo {pages}
  {759} (\bibinfo {year} {1983})}\BibitemShut {NoStop}%
\bibitem [{\citenamefont {Walgraef}(1997)}]{walgraefliv}%
  \BibitemOpen
  \bibfield  {author} {\bibinfo {author} {\bibfnamefont {D.}~\bibnamefont
  {Walgraef}},\ }\href@noop {} {\emph {\bibinfo {title} {Spatio-Temporal
  Pattern Formation: With Examples from Physics, Chemistry, and Materials
  Science}}},\ \bibinfo {edition} {1st}\ ed.,\ Partially Ordered Systems\
  (\bibinfo  {publisher} {Springer-Verlag New York},\ \bibinfo {year}
  {1997})\BibitemShut {NoStop}%
\bibitem [{\citenamefont {Provatas}\ and\ \citenamefont
  {Elder}(2011)}]{provatas2011}%
  \BibitemOpen
  \bibfield  {author} {\bibinfo {author} {\bibfnamefont {N.}~\bibnamefont
  {Provatas}}\ and\ \bibinfo {author} {\bibfnamefont {K.}~\bibnamefont
  {Elder}},\ }\href@noop {} {\emph {\bibinfo {title} {Phase-field methods in
  materials science and engineering}}}\ (\bibinfo  {publisher} {John Wiley \&
  Sons},\ \bibinfo {year} {2011})\BibitemShut {NoStop}%
\bibitem [{\citenamefont {Elder}\ \emph {et~al.}(2007)\citenamefont {Elder},
  \citenamefont {Provatas}, \citenamefont {Berry}, \citenamefont {Stefanovic},\
  and\ \citenamefont {Grant}}]{Provatas1}%
  \BibitemOpen
  \bibfield  {author} {\bibinfo {author} {\bibfnamefont {K.~R.}\ \bibnamefont
  {Elder}}, \bibinfo {author} {\bibfnamefont {N.}~\bibnamefont {Provatas}},
  \bibinfo {author} {\bibfnamefont {J.}~\bibnamefont {Berry}}, \bibinfo
  {author} {\bibfnamefont {P.}~\bibnamefont {Stefanovic}}, \ and\ \bibinfo
  {author} {\bibfnamefont {M.}~\bibnamefont {Grant}},\ }\href {\doibase
  10.1103/PhysRevB.75.064107} {\bibfield  {journal} {\bibinfo  {journal} {Phys.
  Rev. B}\ }\textbf {\bibinfo {volume} {75}},\ \bibinfo {pages} {064107}
  (\bibinfo {year} {2007})}\BibitemShut {NoStop}%
\bibitem [{\citenamefont {Stefanovic}\ \emph {et~al.}(2009)\citenamefont
  {Stefanovic}, \citenamefont {Haataja},\ and\ \citenamefont
  {Provatas}}]{Provatas2}%
  \BibitemOpen
  \bibfield  {author} {\bibinfo {author} {\bibfnamefont {P.}~\bibnamefont
  {Stefanovic}}, \bibinfo {author} {\bibfnamefont {M.}~\bibnamefont {Haataja}},
  \ and\ \bibinfo {author} {\bibfnamefont {N.}~\bibnamefont {Provatas}},\
  }\href {\doibase 10.1103/PhysRevE.80.046107} {\bibfield  {journal} {\bibinfo
  {journal} {Phys. Rev. E}\ }\textbf {\bibinfo {volume} {80}},\ \bibinfo
  {pages} {046107} (\bibinfo {year} {2009})}\BibitemShut {NoStop}%
\bibitem [{\citenamefont {Provatas}\ \emph {et~al.}(2005)\citenamefont
  {Provatas}, \citenamefont {Greenwood}, \citenamefont {Athreya}, \citenamefont
  {Goldenfeld},\ and\ \citenamefont {Dantzig}}]{provatas2005multiscale}%
  \BibitemOpen
  \bibfield  {author} {\bibinfo {author} {\bibfnamefont {N.}~\bibnamefont
  {Provatas}}, \bibinfo {author} {\bibfnamefont {M.}~\bibnamefont {Greenwood}},
  \bibinfo {author} {\bibfnamefont {B.}~\bibnamefont {Athreya}}, \bibinfo
  {author} {\bibfnamefont {N.}~\bibnamefont {Goldenfeld}}, \ and\ \bibinfo
  {author} {\bibfnamefont {J.}~\bibnamefont {Dantzig}},\ }\href {\doibase
  10.1142/S0217979205032917} {\bibfield  {journal} {\bibinfo  {journal}
  {International Journal of Modern Physics B}\ }\textbf {\bibinfo {volume}
  {19}},\ \bibinfo {pages} {4525} (\bibinfo {year} {2005})}\BibitemShut
  {NoStop}%
\bibitem [{\citenamefont {Elder}\ \emph {et~al.}(2002)\citenamefont {Elder},
  \citenamefont {Katakowski}, \citenamefont {Haataja},\ and\ \citenamefont
  {Grant}}]{elder2002modeling}%
  \BibitemOpen
  \bibfield  {author} {\bibinfo {author} {\bibfnamefont {K.~R.}\ \bibnamefont
  {Elder}}, \bibinfo {author} {\bibfnamefont {M.}~\bibnamefont {Katakowski}},
  \bibinfo {author} {\bibfnamefont {M.}~\bibnamefont {Haataja}}, \ and\
  \bibinfo {author} {\bibfnamefont {M.}~\bibnamefont {Grant}},\ }\href
  {\doibase 10.1103/PhysRevLett.88.245701} {\bibfield  {journal} {\bibinfo
  {journal} {Physical review letters}\ }\textbf {\bibinfo {volume} {88}},\
  \bibinfo {pages} {245701} (\bibinfo {year} {2002})}\BibitemShut {NoStop}%
\bibitem [{\citenamefont {Elder}\ and\ \citenamefont
  {Grant}(2004)}]{elder2004modeling}%
  \BibitemOpen
  \bibfield  {author} {\bibinfo {author} {\bibfnamefont {K.~R.}\ \bibnamefont
  {Elder}}\ and\ \bibinfo {author} {\bibfnamefont {M.}~\bibnamefont {Grant}},\
  }\href {\doibase 10.1103/PhysRevE.70.051605} {\bibfield  {journal} {\bibinfo
  {journal} {Physical Review E}\ }\textbf {\bibinfo {volume} {70}},\ \bibinfo
  {pages} {051605} (\bibinfo {year} {2004})}\BibitemShut {NoStop}%
\bibitem [{\citenamefont {Sakaguchi}\ and\ \citenamefont
  {Brand}(1996)}]{sakaguchi1996}%
  \BibitemOpen
  \bibfield  {author} {\bibinfo {author} {\bibfnamefont {H.}~\bibnamefont
  {Sakaguchi}}\ and\ \bibinfo {author} {\bibfnamefont {H.~R.}\ \bibnamefont
  {Brand}},\ }\href {\doibase 10.1016/0167-2789(96)00077-2} {\bibfield
  {journal} {\bibinfo  {journal} {Physica D}\ }\textbf {\bibinfo {volume}
  {97}},\ \bibinfo {pages} {274} (\bibinfo {year} {1996})}\BibitemShut
  {NoStop}%
\bibitem [{\citenamefont {Burke}\ and\ \citenamefont
  {Knobloch}(2006)}]{burke2006localized}%
  \BibitemOpen
  \bibfield  {author} {\bibinfo {author} {\bibfnamefont {J.}~\bibnamefont
  {Burke}}\ and\ \bibinfo {author} {\bibfnamefont {E.}~\bibnamefont
  {Knobloch}},\ }\href {\doibase 10.1103/PhysRevE.73.056211} {\bibfield
  {journal} {\bibinfo  {journal} {Physical Review E}\ }\textbf {\bibinfo
  {volume} {73}},\ \bibinfo {pages} {056211} (\bibinfo {year}
  {2006})}\BibitemShut {NoStop}%
\bibitem [{\citenamefont {Vitral}\ \emph {et~al.}(2019)\citenamefont {Vitral},
  \citenamefont {Leo},\ and\ \citenamefont {Vi{\~n}als}}]{vitral2019role}%
  \BibitemOpen
  \bibfield  {author} {\bibinfo {author} {\bibfnamefont {E.}~\bibnamefont
  {Vitral}}, \bibinfo {author} {\bibfnamefont {P.~H.}\ \bibnamefont {Leo}}, \
  and\ \bibinfo {author} {\bibfnamefont {J.}~\bibnamefont {Vi{\~n}als}},\
  }\href {\doibase 10.1103/PhysRevE.100.032805} {\bibfield  {journal} {\bibinfo
   {journal} {Physical Review E}\ }\textbf {\bibinfo {volume} {100}},\ \bibinfo
  {pages} {032805} (\bibinfo {year} {2019})}\BibitemShut {NoStop}%
\bibitem [{\citenamefont {Walton}(1983)}]{walton1983onset}%
  \BibitemOpen
  \bibfield  {author} {\bibinfo {author} {\bibfnamefont {I.}~\bibnamefont
  {Walton}},\ }\href {\doibase 10.1017/S002211208300140} {\bibfield  {journal}
  {\bibinfo  {journal} {Journal of Fluid Mechanics}\ }\textbf {\bibinfo
  {volume} {131}},\ \bibinfo {pages} {455} (\bibinfo {year}
  {1983})}\BibitemShut {NoStop}%
\bibitem [{\citenamefont {Cross}(1982{\natexlab{a}})}]{cross1982boundary}%
  \BibitemOpen
  \bibfield  {author} {\bibinfo {author} {\bibfnamefont {M.}~\bibnamefont
  {Cross}},\ }\href {\doibase 10.1063/1.863835} {\bibfield  {journal} {\bibinfo
   {journal} {The Physics of Fluids}\ }\textbf {\bibinfo {volume} {25}},\
  \bibinfo {pages} {936} (\bibinfo {year} {1982}{\natexlab{a}})}\BibitemShut
  {NoStop}%
\bibitem [{\citenamefont {Cross}(1982{\natexlab{b}})}]{cross1982ingredients}%
  \BibitemOpen
  \bibfield  {author} {\bibinfo {author} {\bibfnamefont {M.}~\bibnamefont
  {Cross}},\ }\href {\doibase 10.1103/PhysRevA.25.1065} {\bibfield  {journal}
  {\bibinfo  {journal} {Physical Review A}\ }\textbf {\bibinfo {volume} {25}},\
  \bibinfo {pages} {1065} (\bibinfo {year} {1982}{\natexlab{b}})}\BibitemShut
  {NoStop}%
\bibitem [{\citenamefont {Cross}\ and\ \citenamefont
  {Hohenberg}(1993)}]{cross1993}%
  \BibitemOpen
  \bibfield  {author} {\bibinfo {author} {\bibfnamefont {M.~C.}\ \bibnamefont
  {Cross}}\ and\ \bibinfo {author} {\bibfnamefont {P.~C.}\ \bibnamefont
  {Hohenberg}},\ }\href {\doibase 10.1103/RevModPhys.65.851} {\bibfield
  {journal} {\bibinfo  {journal} {Rev. Modern Phys.}\ }\textbf {\bibinfo
  {volume} {65}},\ \bibinfo {pages} {851} (\bibinfo {year} {1993})}\BibitemShut
  {NoStop}%
\bibitem [{\citenamefont {Greenside}\ and\ \citenamefont
  {Coughran}(1984)}]{Greenside1984}%
  \BibitemOpen
  \bibfield  {author} {\bibinfo {author} {\bibfnamefont {H.~S.}\ \bibnamefont
  {Greenside}}\ and\ \bibinfo {author} {\bibfnamefont {W.~M.}\ \bibnamefont
  {Coughran}},\ }\href {\doibase 10.1103/PhysRevA.30.398} {\bibfield  {journal}
  {\bibinfo  {journal} {Physical Review A}\ }\textbf {\bibinfo {volume} {30}},\
  \bibinfo {pages} {398} (\bibinfo {year} {1984})}\BibitemShut {NoStop}%
\bibitem [{\citenamefont {Manneville}(1995)}]{manneville1995dissipative}%
  \BibitemOpen
  \bibfield  {author} {\bibinfo {author} {\bibfnamefont {P.}~\bibnamefont
  {Manneville}},\ }\href@noop {} {\emph {\bibinfo {title} {Dissipative
  structures and weak turbulence}}}\ (\bibinfo  {publisher} {Springer},\
  \bibinfo {year} {1995})\BibitemShut {NoStop}%
\bibitem [{\citenamefont {Sruljes}(1970)}]{sruljes1979}%
  \BibitemOpen
  \bibfield  {author} {\bibinfo {author} {\bibfnamefont {J.~A.}\ \bibnamefont
  {Sruljes}},\ }\emph {\bibinfo {title} {Zellularkonvection in Beh\"altern mit
  Horizontalen Temperauregradienten}},\ \href@noop {} {Ph.D. thesis},\ \bibinfo
   {school} {Fakult\"at f\"ur {Machinenbau}, Univ. Karlsruhe}, \bibinfo
  {address} {Karlsruhe} (\bibinfo {year} {1970})\BibitemShut {NoStop}%
\bibitem [{\citenamefont {Pontes}(1994)}]{pontes1994}%
  \BibitemOpen
  \bibfield  {author} {\bibinfo {author} {\bibfnamefont {J.}~\bibnamefont
  {Pontes}},\ }\emph {\bibinfo {title} {Pattern formation in spatially ramped
  {R}ayleigh-{B}{\'e}nard systems}},\ \href@noop {} {Ph.D. thesis},\ \bibinfo
  {school} {Free University of Brussels}, \bibinfo {address} {Brussels,
  Belgium} (\bibinfo {year} {1994})\BibitemShut {NoStop}%
\bibitem [{\citenamefont {Pontes}\ \emph {et~al.}(2008)\citenamefont {Pontes},
  \citenamefont {Walgraef},\ and\ \citenamefont {Christov}}]{pontes2008}%
  \BibitemOpen
  \bibfield  {author} {\bibinfo {author} {\bibfnamefont {J.}~\bibnamefont
  {Pontes}}, \bibinfo {author} {\bibfnamefont {D.}~\bibnamefont {Walgraef}}, \
  and\ \bibinfo {author} {\bibfnamefont {C.~I.}\ \bibnamefont {Christov}},\
  }\href {\doibase 10.6062/jcis.2008.01.01.0002} {\bibfield  {journal}
  {\bibinfo  {journal} {Journal of Computational Interdisciplinary Sciences}\
  }\textbf {\bibinfo {volume} {1}},\ \bibinfo {pages} {11} (\bibinfo {year}
  {2008})}\BibitemShut {NoStop}%
\bibitem [{\citenamefont {Hilali}\ \emph {et~al.}(1995)\citenamefont {Hilali},
  \citenamefont {M\'etens}, \citenamefont {Borckmans},\ and\ \citenamefont
  {Dewel}}]{hilali1995}%
  \BibitemOpen
  \bibfield  {author} {\bibinfo {author} {\bibfnamefont {M.~F.}\ \bibnamefont
  {Hilali}}, \bibinfo {author} {\bibfnamefont {S.}~\bibnamefont {M\'etens}},
  \bibinfo {author} {\bibfnamefont {P.}~\bibnamefont {Borckmans}}, \ and\
  \bibinfo {author} {\bibfnamefont {G.}~\bibnamefont {Dewel}},\ }\href
  {\doibase 10.1103/PhysRevE.51.2046} {\bibfield  {journal} {\bibinfo
  {journal} {Physical Review E}\ }\textbf {\bibinfo {volume} {51}},\ \bibinfo
  {pages} {2046} (\bibinfo {year} {1995})}\BibitemShut {NoStop}%
\bibitem [{\citenamefont {Malomed}\ and\ \citenamefont
  {Nepomnyashchy}(1993)}]{Malomed_1993}%
  \BibitemOpen
  \bibfield  {author} {\bibinfo {author} {\bibfnamefont {B.~A.}\ \bibnamefont
  {Malomed}}\ and\ \bibinfo {author} {\bibfnamefont {A.~A.}\ \bibnamefont
  {Nepomnyashchy}},\ }\href {\doibase 10.1209/0295-5075/21/2/013} {\bibfield
  {journal} {\bibinfo  {journal} {Europhysics Letters ({EPL})}\ }\textbf
  {\bibinfo {volume} {21}},\ \bibinfo {pages} {195} (\bibinfo {year}
  {1993})}\BibitemShut {NoStop}%
\bibitem [{\citenamefont {Morris}\ \emph {et~al.}(1993)\citenamefont {Morris},
  \citenamefont {Bodenschatz}, \citenamefont {Cannell},\ and\ \citenamefont
  {Ahlers}}]{morris1993spiral}%
  \BibitemOpen
  \bibfield  {author} {\bibinfo {author} {\bibfnamefont {S.~W.}\ \bibnamefont
  {Morris}}, \bibinfo {author} {\bibfnamefont {E.}~\bibnamefont {Bodenschatz}},
  \bibinfo {author} {\bibfnamefont {D.~S.}\ \bibnamefont {Cannell}}, \ and\
  \bibinfo {author} {\bibfnamefont {G.}~\bibnamefont {Ahlers}},\ }\href@noop {}
  {\bibfield  {journal} {\bibinfo  {journal} {Physical review letters}\
  }\textbf {\bibinfo {volume} {71}},\ \bibinfo {pages} {2026} (\bibinfo {year}
  {1993})}\BibitemShut {NoStop}%
\bibitem [{\citenamefont {Vitral}\ \emph {et~al.}(2020)\citenamefont {Vitral},
  \citenamefont {Mukherjee}, \citenamefont {Leo}, \citenamefont {Vi{\~n}als},
  \citenamefont {Paul},\ and\ \citenamefont {Huang}}]{vitral2020spiral}%
  \BibitemOpen
  \bibfield  {author} {\bibinfo {author} {\bibfnamefont {E.}~\bibnamefont
  {Vitral}}, \bibinfo {author} {\bibfnamefont {S.}~\bibnamefont {Mukherjee}},
  \bibinfo {author} {\bibfnamefont {P.~H.}\ \bibnamefont {Leo}}, \bibinfo
  {author} {\bibfnamefont {J.}~\bibnamefont {Vi{\~n}als}}, \bibinfo {author}
  {\bibfnamefont {M.~R.}\ \bibnamefont {Paul}}, \ and\ \bibinfo {author}
  {\bibfnamefont {Z.-F.}\ \bibnamefont {Huang}},\ }\href@noop {} {\bibfield
  {journal} {\bibinfo  {journal} {Physical Review Fluids}\ }\textbf {\bibinfo
  {volume} {5}},\ \bibinfo {pages} {093501} (\bibinfo {year}
  {2020})}\BibitemShut {NoStop}%
\bibitem [{\citenamefont {Huang}\ and\ \citenamefont
  {Vi{\~n}als}(2004)}]{huang2004shear}%
  \BibitemOpen
  \bibfield  {author} {\bibinfo {author} {\bibfnamefont {Z.-F.}\ \bibnamefont
  {Huang}}\ and\ \bibinfo {author} {\bibfnamefont {J.}~\bibnamefont
  {Vi{\~n}als}},\ }\href@noop {} {\bibfield  {journal} {\bibinfo  {journal}
  {Physical Review E}\ }\textbf {\bibinfo {volume} {69}},\ \bibinfo {pages}
  {041504} (\bibinfo {year} {2004})}\BibitemShut {NoStop}%
\bibitem [{\citenamefont {Vitral}\ \emph {et~al.}(2021)\citenamefont {Vitral},
  \citenamefont {Leo},\ and\ \citenamefont {Vi{\~n}als}}]{vitral2021phase}%
  \BibitemOpen
  \bibfield  {author} {\bibinfo {author} {\bibfnamefont {E.}~\bibnamefont
  {Vitral}}, \bibinfo {author} {\bibfnamefont {P.~H.}\ \bibnamefont {Leo}}, \
  and\ \bibinfo {author} {\bibfnamefont {J.}~\bibnamefont {Vi{\~n}als}},\
  }\href@noop {} {\bibfield  {journal} {\bibinfo  {journal} {Soft Matter}\
  }\textbf {\bibinfo {volume} {17}},\ \bibinfo {pages} {6140} (\bibinfo {year}
  {2021})}\BibitemShut {NoStop}%
\bibitem [{\citenamefont {Hiscock}\ and\ \citenamefont
  {Megason}(2015)}]{hiscock2015orientation}%
  \BibitemOpen
  \bibfield  {author} {\bibinfo {author} {\bibfnamefont {T.~W.}\ \bibnamefont
  {Hiscock}}\ and\ \bibinfo {author} {\bibfnamefont {S.~G.}\ \bibnamefont
  {Megason}},\ }\href {\doibase 10.1016/j.cels.2015.12.001} {\bibfield
  {journal} {\bibinfo  {journal} {Cell systems}\ }\textbf {\bibinfo {volume}
  {1}},\ \bibinfo {pages} {408} (\bibinfo {year} {2015})}\BibitemShut {NoStop}%
\bibitem [{\citenamefont {Rapp}\ \emph {et~al.}(2016)\citenamefont {Rapp},
  \citenamefont {Bergmann},\ and\ \citenamefont
  {Zimmermann}}]{rapp2016pattern}%
  \BibitemOpen
  \bibfield  {author} {\bibinfo {author} {\bibfnamefont {L.}~\bibnamefont
  {Rapp}}, \bibinfo {author} {\bibfnamefont {F.}~\bibnamefont {Bergmann}}, \
  and\ \bibinfo {author} {\bibfnamefont {W.}~\bibnamefont {Zimmermann}},\
  }\href@noop {} {\bibfield  {journal} {\bibinfo  {journal} {EPL (Europhysics
  Letters)}\ }\textbf {\bibinfo {volume} {113}},\ \bibinfo {pages} {28006}
  (\bibinfo {year} {2016})}\BibitemShut {NoStop}%
\bibitem [{\citenamefont {Kaoui}\ \emph {et~al.}(2015)\citenamefont {Kaoui},
  \citenamefont {Guckenberger}, \citenamefont {Krekhov}, \citenamefont
  {Ziebert},\ and\ \citenamefont {Zimmermann}}]{kaoui2015coexistence}%
  \BibitemOpen
  \bibfield  {author} {\bibinfo {author} {\bibfnamefont {B.}~\bibnamefont
  {Kaoui}}, \bibinfo {author} {\bibfnamefont {A.}~\bibnamefont {Guckenberger}},
  \bibinfo {author} {\bibfnamefont {A.}~\bibnamefont {Krekhov}}, \bibinfo
  {author} {\bibfnamefont {F.}~\bibnamefont {Ziebert}}, \ and\ \bibinfo
  {author} {\bibfnamefont {W.}~\bibnamefont {Zimmermann}},\ }\href@noop {}
  {\bibfield  {journal} {\bibinfo  {journal} {New Journal of Physics}\ }\textbf
  {\bibinfo {volume} {17}},\ \bibinfo {pages} {103015} (\bibinfo {year}
  {2015})}\BibitemShut {NoStop}%
\bibitem [{\citenamefont {Morgan}\ and\ \citenamefont
  {Dawes}(2014)}]{morgan2014swift}%
  \BibitemOpen
  \bibfield  {author} {\bibinfo {author} {\bibfnamefont {D.}~\bibnamefont
  {Morgan}}\ and\ \bibinfo {author} {\bibfnamefont {J.~H.}\ \bibnamefont
  {Dawes}},\ }\href {\doibase 10.1016/j.physd.2013.11.018} {\bibfield
  {journal} {\bibinfo  {journal} {Physica D}\ }\textbf {\bibinfo {volume}
  {270}},\ \bibinfo {pages} {60} (\bibinfo {year} {2014})}\BibitemShut
  {NoStop}%
\bibitem [{\citenamefont {Walton}(1982)}]{walton1982onset}%
  \BibitemOpen
  \bibfield  {author} {\bibinfo {author} {\bibfnamefont {I.}~\bibnamefont
  {Walton}},\ }\href {\doibase 10.1002/sapm1982673199} {\bibfield  {journal}
  {\bibinfo  {journal} {Studies in Applied Mathematics}\ }\textbf {\bibinfo
  {volume} {67}},\ \bibinfo {pages} {199} (\bibinfo {year} {1982})}\BibitemShut
  {NoStop}%
\bibitem [{\citenamefont {Coelho}\ \emph {et~al.}(2020)\citenamefont {Coelho},
  \citenamefont {Vitral}, \citenamefont {Pontes},\ and\ \citenamefont
  {Mangiavacchi}}]{coelho2020}%
  \BibitemOpen
  \bibfield  {author} {\bibinfo {author} {\bibfnamefont {D.~L.}\ \bibnamefont
  {Coelho}}, \bibinfo {author} {\bibfnamefont {E.}~\bibnamefont {Vitral}},
  \bibinfo {author} {\bibfnamefont {J.}~\bibnamefont {Pontes}}, \ and\ \bibinfo
  {author} {\bibfnamefont {N.}~\bibnamefont {Mangiavacchi}},\ }\href@noop {}
  {\enquote {\bibinfo {title} {Numerical scheme for solving the nonuniformly
  forced cubic and quintic {S}wift-{H}ohenberg equations strictly respecting
  the {L}yapunov functional},}\ } (\bibinfo {year} {2020}),\ \Eprint
  {http://arxiv.org/abs/2007.16080} {arXiv:2007.16080 [nlin.PS]} \BibitemShut
  {NoStop}%
\bibitem [{\citenamefont {Christov}\ and\ \citenamefont
  {Pontes}(2002)}]{C.I-2002}%
  \BibitemOpen
  \bibfield  {author} {\bibinfo {author} {\bibfnamefont {C.}~\bibnamefont
  {Christov}}\ and\ \bibinfo {author} {\bibfnamefont {J.}~\bibnamefont
  {Pontes}},\ }\href {\doibase 10.1016/S0895-7177(01)00151-0} {\bibfield
  {journal} {\bibinfo  {journal} {Mathematical and Computer Modelling}\
  }\textbf {\bibinfo {volume} {35}} (\bibinfo {year} {2002}),\
  10.1016/S0895-7177(01)00151-0}\BibitemShut {NoStop}%
\bibitem [{\citenamefont {Segel}(1969{\natexlab{a}})}]{segel1969distant}%
  \BibitemOpen
  \bibfield  {author} {\bibinfo {author} {\bibfnamefont {L.~A.}\ \bibnamefont
  {Segel}},\ }\href@noop {} {\bibfield  {journal} {\bibinfo  {journal} {Journal
  of Fluid Mechanics}\ }\textbf {\bibinfo {volume} {38}},\ \bibinfo {pages}
  {203} (\bibinfo {year} {1969}{\natexlab{a}})}\BibitemShut {NoStop}%
\bibitem [{\citenamefont {Daniels}(1978)}]{daniels1978effect}%
  \BibitemOpen
  \bibfield  {author} {\bibinfo {author} {\bibfnamefont {P.}~\bibnamefont
  {Daniels}},\ }\href@noop {} {\bibfield  {journal} {\bibinfo  {journal}
  {Proceedings of the Royal Society of London. A. Mathematical and Physical
  Sciences}\ }\textbf {\bibinfo {volume} {358}},\ \bibinfo {pages} {173}
  (\bibinfo {year} {1978})}\BibitemShut {NoStop}%
\bibitem [{\citenamefont {Cross}\ \emph {et~al.}(1980)\citenamefont {Cross},
  \citenamefont {Daniels}, \citenamefont {Hohenberg},\ and\ \citenamefont
  {Siggia}}]{cross1980effect}%
  \BibitemOpen
  \bibfield  {author} {\bibinfo {author} {\bibfnamefont {M.}~\bibnamefont
  {Cross}}, \bibinfo {author} {\bibfnamefont {P.}~\bibnamefont {Daniels}},
  \bibinfo {author} {\bibfnamefont {P.}~\bibnamefont {Hohenberg}}, \ and\
  \bibinfo {author} {\bibfnamefont {E.}~\bibnamefont {Siggia}},\ }\href@noop {}
  {\bibfield  {journal} {\bibinfo  {journal} {Physical Review Letters}\
  }\textbf {\bibinfo {volume} {45}},\ \bibinfo {pages} {898} (\bibinfo {year}
  {1980})}\BibitemShut {NoStop}%
\bibitem [{\citenamefont {Hoyle}(2006)}]{hoyle2006pattern}%
  \BibitemOpen
  \bibfield  {author} {\bibinfo {author} {\bibfnamefont {R.~B.}\ \bibnamefont
  {Hoyle}},\ }\href@noop {} {\emph {\bibinfo {title} {Pattern formation: an
  introduction to methods}}}\ (\bibinfo  {publisher} {Cambridge University
  Press},\ \bibinfo {year} {2006})\BibitemShut {NoStop}%
\bibitem [{\citenamefont {Bestehorn}\ and\ \citenamefont
  {P{\'e}rez-Garc{\'\i}a}(1992)}]{bestehorn1992study}%
  \BibitemOpen
  \bibfield  {author} {\bibinfo {author} {\bibfnamefont {M.}~\bibnamefont
  {Bestehorn}}\ and\ \bibinfo {author} {\bibfnamefont {C.}~\bibnamefont
  {P{\'e}rez-Garc{\'\i}a}},\ }\href@noop {} {\bibfield  {journal} {\bibinfo
  {journal} {Physica D: Nonlinear Phenomena}\ }\textbf {\bibinfo {volume}
  {61}},\ \bibinfo {pages} {67} (\bibinfo {year} {1992})}\BibitemShut {NoStop}%
\bibitem [{\citenamefont {Christov}\ \emph {et~al.}(1997)\citenamefont
  {Christov}, \citenamefont {Pontes}, \citenamefont {Walgraef},\ and\
  \citenamefont {Velarde}}]{C.I-1997}%
  \BibitemOpen
  \bibfield  {author} {\bibinfo {author} {\bibfnamefont {C.}~\bibnamefont
  {Christov}}, \bibinfo {author} {\bibfnamefont {J.}~\bibnamefont {Pontes}},
  \bibinfo {author} {\bibfnamefont {D.}~\bibnamefont {Walgraef}}, \ and\
  \bibinfo {author} {\bibfnamefont {M.}~\bibnamefont {Velarde}},\ }\href
  {\doibase 10.1016/S0045-7825(96)01176-0} {\bibfield  {journal} {\bibinfo
  {journal} {Computer Methods in Applied Mechanics and Engineering}\ }\textbf
  {\bibinfo {volume} {148}} (\bibinfo {year} {1997}),\
  10.1016/S0045-7825(96)01176-0}\BibitemShut {NoStop}%
\bibitem [{\citenamefont {Vitral}\ \emph {et~al.}(2018)\citenamefont {Vitral},
  \citenamefont {Walgraef}, \citenamefont {Pontes}, \citenamefont {Anjos},\
  and\ \citenamefont {Mangiavacchi}}]{Vitral-2018}%
  \BibitemOpen
  \bibfield  {author} {\bibinfo {author} {\bibfnamefont {E.}~\bibnamefont
  {Vitral}}, \bibinfo {author} {\bibfnamefont {D.}~\bibnamefont {Walgraef}},
  \bibinfo {author} {\bibfnamefont {J.}~\bibnamefont {Pontes}}, \bibinfo
  {author} {\bibfnamefont {G.}~\bibnamefont {Anjos}}, \ and\ \bibinfo {author}
  {\bibfnamefont {N.}~\bibnamefont {Mangiavacchi}},\ }\href {\doibase
  10.1016/j.commatsci.2018.01.034} {\bibfield  {journal} {\bibinfo  {journal}
  {Computational Materials Science}\ }\textbf {\bibinfo {volume} {146}}
  (\bibinfo {year} {2018}),\ 10.1016/j.commatsci.2018.01.034}\BibitemShut
  {NoStop}%
\bibitem [{\citenamefont {Newell}\ and\ \citenamefont
  {Whitehead}(1969)}]{newell1969}%
  \BibitemOpen
  \bibfield  {author} {\bibinfo {author} {\bibfnamefont {A.~C.}\ \bibnamefont
  {Newell}}\ and\ \bibinfo {author} {\bibfnamefont {J.~A.}\ \bibnamefont
  {Whitehead}},\ }\href {\doibase 10.1017/S0022112069000176} {\bibfield
  {journal} {\bibinfo  {journal} {J. Fluid Mech.}\ }\textbf {\bibinfo {volume}
  {38}},\ \bibinfo {pages} {279} (\bibinfo {year} {1969})}\BibitemShut
  {NoStop}%
\bibitem [{\citenamefont {Segel}(1969{\natexlab{b}})}]{Segel1969}%
  \BibitemOpen
  \bibfield  {author} {\bibinfo {author} {\bibfnamefont {L.~A.}\ \bibnamefont
  {Segel}},\ }\href {\doibase 10.1017/S0022112069000127} {\bibfield  {journal}
  {\bibinfo  {journal} {Journal of Fluid Mechanics}\ }\textbf {\bibinfo
  {volume} {38:1}},\ \bibinfo {pages} {203} (\bibinfo {year}
  {1969}{\natexlab{b}})}\BibitemShut {NoStop}%
\bibitem [{\citenamefont {Ruppert}\ \emph {et~al.}(2020)\citenamefont
  {Ruppert}, \citenamefont {Ziebert},\ and\ \citenamefont
  {Zimmermann}}]{ruppert2020nonlinear}%
  \BibitemOpen
  \bibfield  {author} {\bibinfo {author} {\bibfnamefont {M.}~\bibnamefont
  {Ruppert}}, \bibinfo {author} {\bibfnamefont {F.}~\bibnamefont {Ziebert}}, \
  and\ \bibinfo {author} {\bibfnamefont {W.}~\bibnamefont {Zimmermann}},\
  }\href {\doibase 10.1088/1367-2630/ab7f92} {\bibfield  {journal} {\bibinfo
  {journal} {New Journal of Physics}\ }\textbf {\bibinfo {volume} {22}},\
  \bibinfo {pages} {052001} (\bibinfo {year} {2020})}\BibitemShut {NoStop}%
\bibitem [{\citenamefont {Auzerais}\ \emph {et~al.}(2016)\citenamefont
  {Auzerais}, \citenamefont {Jarno}, \citenamefont {Ezersky},\ and\
  \citenamefont {Marin}}]{auzerais2016formation}%
  \BibitemOpen
  \bibfield  {author} {\bibinfo {author} {\bibfnamefont {A.}~\bibnamefont
  {Auzerais}}, \bibinfo {author} {\bibfnamefont {A.}~\bibnamefont {Jarno}},
  \bibinfo {author} {\bibfnamefont {A.}~\bibnamefont {Ezersky}}, \ and\
  \bibinfo {author} {\bibfnamefont {F.}~\bibnamefont {Marin}},\ }\href
  {\doibase 10.1103/PhysRevE.94.052903} {\bibfield  {journal} {\bibinfo
  {journal} {Physical Review E}\ }\textbf {\bibinfo {volume} {94}},\ \bibinfo
  {pages} {052903} (\bibinfo {year} {2016})}\BibitemShut {NoStop}%
\bibitem [{\citenamefont {Boyd}(2001)}]{boyd2001chebyshev}%
  \BibitemOpen
  \bibfield  {author} {\bibinfo {author} {\bibfnamefont {J.~P.}\ \bibnamefont
  {Boyd}},\ }\href@noop {} {\emph {\bibinfo {title} {Chebyshev and Fourier
  spectral methods}}}\ (\bibinfo  {publisher} {Courier Corporation},\ \bibinfo
  {year} {2001})\BibitemShut {NoStop}%
\end{thebibliography}%

\end{document}